\documentclass[preprint,12pt,authoryear]{elsarticle}

\usepackage{amssymb}
\usepackage{amsmath}

\usepackage{hyperref}
\usepackage{wrapfig}

\journal{Nuclear Physics B}

\begin{document}

\begin{frontmatter}



\title{Body Ownership Affects the Processing of Sensorimotor Contingencies in Virtual Reality} 


\author{Evan G. Center\fnref{equal}} 

\affiliation{organization={University of Oulu},
            addressline={Center for Ubiquitous Computing},
            country={Finland}}

\author{Matti Pouke\fnref{equal}} 

\affiliation{organization={University of Oulu},
            addressline={Center for Ubiquitous Computing},
            country={Finland}}

\author{Alessandro Nardi} 

\affiliation{organization={University of Oulu},
            addressline={Center for Ubiquitous Computing},
            country={Finland}}

\author{Lukas Gehrke} 

\affiliation{organization={Technische Universität Berlin},
            addressline={Berlin Mobile Brain/Body Imaging lab (BeMoBIL)},
            country={Germany}}

\author{Klaus Gramann} 

\affiliation{organization={Technische Universität Berlin},
            addressline={Berlin Mobile Brain/Body Imaging lab (BeMoBIL)},
            country={Germany}}

\author{Timo Ojala} 

\affiliation{organization={University of Oulu},
            addressline={Center for Ubiquitous Computing},
            country={Finland}}

\author{Steven M. LaValle} 

\affiliation{organization={University of Oulu},
            addressline={Center for Ubiquitous Computing},
            country={Finland}}

\fntext[equal]{These authors contributed equally to this work.}

\begin{abstract}

Presence in virtual reality (VR), the subjective sense of "being there" in a virtual environment, is notoriously difficult to measure. Electroencephalography (EEG) may offer a promising, unobtrusive means of assessing a user’s momentary state of presence. Unlike traditional questionnaires, EEG does not interrupt the experience or rely on users’ retrospective self-reports, thereby avoiding interference with the very state it aims to capture. Previous research has attempted to quantify presence in virtual environments using event-related potentials (ERPs). We contend, however, that previous efforts have fallen short of fully realizing this goal, failing to either A) independently manipulate presence, B) validate their measure of presence against traditional techniques, C) adequately separate the constructs of presence and attention, and/or D) implement a realistic and immersive environment and task. We address these shortcomings in a preregistered ERP experiment in which participants play an engaging target shooting game in VR. ERPs are time-locked to the release of a ball from a sling. We induce breaks in presence (BIPs) by freezing the ball's release on a minority of trials. Embodiment is manipulated by allowing manual manipulation of the sling with a realistic avatar in one condition (embodied condition) and passive manipulation with only controllers in another (non-embodied condition). We support our predictions that the N2, the P3b, and the N400, are selectively sensitive towards specific components of these manipulations. The pattern of findings carries significant implications for theories of presence, which have been seldom addressed in previous ERP investigations on this topic. 
\end{abstract}



\begin{keyword}
virtual reality \sep place illusion \sep plausibility illusion \sep breaks-in-presence \sep embodiment \sep electroencephalography \sep event-related potentials



\end{keyword}

\end{frontmatter}



\section{Introduction}
\label{sec1}
Virtual reality often compels users to respond physiologically and consciously as if virtual events were real, otherwise known as the experience of being present. The experience of presence is thought to arise from two distinct illusions: Place Illusion (PI), the feeling of being located inside the virtual environment, and Plausibility Illusion (PsI), the feeling that the depicted events are actually occurring.

Despite compelling theoretical accounts that PI is driven by bottom-up sensorimotor contingencies and PsI by top-down environmental `coherence', the field lacks moment-to-moment neural evidence to index and begin to dissociate these illusions. Standard post-experience questionnaires are too coarse and miss the transient `breaks in presence' (BIPs) that occur when either illusion collapses.

In this report, we asked whether specific event-related potentials (ERPs) could be used to identify PI-related and PsI-related processing during controlled BIPs in an immersive VR task.

We designed a VR slingshot game in which brief freezes of the virtual hand and/or slingshot were used to elicit BIPs under two conditions: with and without a first-person virtual body. Analyzing event-related activity in the electroencephalography (EEG) signal, we tested several preregistered hypotheses on established ERPs. First, does the N2 index PI-related sensorimotor mismatch, second, does the N400 reflect PsI-related semantic violations, and lastly, does the P3b capture embodiment-driven context updating? The results are discussed within the context of existing theories of presence to advance the neural quantification of presence-based illusions.

\section{Related Work}

Although the exact definition of presence sometimes varies within the VR literature, one of the more commonly accepted models is Slater's view that presence consists of two orthogonal illusions called Place Illusion (PI) and Plausibility Illusion (PsI) \citep{slater2009place,slater2022separate}. According to this model, PI is the illusion of being in another location, yet having the sure knowledge one is not (this sensation is sometimes referred to as \textit{Spatial Presence} in literature; \citealt{skarbez2017survey}). PsI, on the other hand, is the illusion of events in the VR experience actually happening, yet having the sure knowledge that they are not. It is argued that these two illusions are orthogonal; for example, one can experience PsI without PI by watching TV \citep{banos2000presence,skarbez2017survey}, or by mistaking a physical mannequin for a real person \citep{slater2010simulating}, and that the user responds realistically to VR events specifically when both of these illusions are taking place \citep{slater2009place}. Disentangling these illusions through empirical behavioral studies has proven to be nontrivial \citep{slater2022separate}. However, if these illusions are truly orthogonal, we should be able to identify dissociable brain processes related to each. We propose that neuroimaging techniques, such as EEG, could help overcoming earlier limitations of behavioral studies in this respect.

Multiple cognitive and neuroscientific models of presence have been discussed by the VR community, however, the amount of empirical evidence backing each model is so far limited. \cite{gonzalez2017model} proposed a model according to which VR illusions are sustained by the interplay of bottom-up multisensory processing, sensorimotor self-awareness frameworks, and top-down predictions, especially when the stimuli can be reconciled without semantic violations. Regarding PI, and PsI, specifically, \cite{hofer2020role} suggested that PI is a bottom-up process whereas PsI is a top-down one. \cite{latoschikcongruence} suggest that different VR illusions are elicited by weighted congruence activations of stimuli that can be categorized into three layers called \textit{sensation, perception}, and \textit{cognition}. Perhaps the most well-known model is the so-called Slater/Skarbez model according to which PI is driven by the sensorimotor contingencies (SMCs) provided by the VR system whereas PsI is enabled by \textit{coherence} -- the extent to which the high-level content of the virtual environment matches the user's predictions \citep{skarbez2017survey}. Although not specifically mentioned in the original \cite{skarbez2017survey} article, this division also suggests that PI would be more driven by earlier perceptual processes whereas PsI would be the result of later cognitive ones. This dual-nature approach for presence is well-known in the VR research community, however, it has attracted relatively little empirical investigation from neuroscience. 

\subsection{Designing for PI and PsI}
Much of presence research is devoted to understanding which factors contribute to presence or alternatively break it. Although in colloquial speech, \textit{immersion} has a role similar to presence in media experiences (e.g., \textit{``I was immersed in the story throughout the book''} or \textit{``the weird behavior of the characters broke my immersion in this video game''}), in the context of VR research, immersion usually refers to the set of hardware and software factors that contribute to presence \citep{skarbez2017survey,slater1997framework}. Interestingly, the aesthetic quality of graphics appears to have little effect on presence. For example, an early study by \cite{zimmons2003influence} found that even an untextured black-and-white grid representation of a virtual pit elicited a fear response in participants and increases in graphics fidelity elicited no measurable differences. According to Slater, immersion equates to the extent of the \textit{sensorimotor contingencies} (SMCs) that the VR system can support \citep{slater2009place,slater2022separate}. 

Understanding the brain in a predictive coding framework \citep{Clark2013-ah, Friston2010-hy, Rao1999-zr, Seth2013-jl} can explain why SMCs and coherence matter for the presence experience from a neuroscientific perspective. According to hierarchical predictive coding, the brain constructs models that continuously predict the sensory consequences of (top-down) self-generated action and minimise (bottom-up) precision-weighted prediction error. Here, SMCs constitute the learned mappings between actions and their predicted sensory effects which reflect the same core principles emphasized in theories of embodied cognition, which view perception and action as inseparably grounded in the body's capabilites to interact with its environment (in other words, an \textit{affordance}; \citealt{gibson2014ecological,noe2004action}).

When a VR system correctly supports these contingencies, such as when a motor command reliably produces the predicted reafferent input, low-level prediction error remains small, and PI is sustained. For example, head turns or hand movements yield congruent visual and haptic feedback. Conversely, a sudden mismatch, such as a tracked hand freezing in mid-air, constitutes a spike in prediction error that the system cannot immediately explain away, manifesting phenomenologically as a BIP. At higher levels of the hierarchy, causal, narrative, and social regularities are encoded, i.e., \textit{coherence}. If virtual agents behave implausibly or objects violate physics, prediction error rises at these levels, jeopardizing PsI. Guided by this framework, our study deliberately perturbed SMCs, briefly freezing the user-controlled hand or slingshot, to induce BIPs and examine how PI and PsI are affected by violations of embodied predictions under different levels of embodiment. Designing VR to minimize prediction error at the intended levels therefore aligns with both SMC-based and predictive coding explanations of presence.

In the context of VR specifically, SMCs refer to the extent to which the VR system can render stimuli according to natural sensorimotor actions such as turning one's head, moving one's arms, and so on. More specifically, if the set of \textit{valid actions}, which is the union of natural sensorimotor actions, matches the effectual actions supported by the VR experience \citep{slater2009place}, PI will be sustained. A break in SMCs occurs when the display of the VR system fails to support these actions. For example, such a break might occur when a CAVE user looks outside of the display coverage, or when a user pushes their hand into an obstacle, but the system doesn't respond with the appropriate tactile or visual feedback. According to a meta-analysis by \cite{cummings2016immersive}, immersion factors such as tracking quality and display properties (i.e., FOV, stereoscopy) contribute specifically towards spatial presence (i.e., PI) more than, for example, graphics or audio fidelity.

In their articles, Skarbez\citep{skarbez2016plausibility,skarbez2017survey,skarbez2020immersion} revisits Slater's model and argues that if immersion is the set of factors contributing towards PI, there must be a similar set of factors contributing towards PsI, as well. Skarbez suggests the concept of \textit{coherence}, which refers to the extent the VR experience behaves in a reasonable or expected way. The difference between immersion and coherence is, however, that although the immersion of a VR system can be defined in an objective manner, the same is not true for coherence, as expectations are always subjective. Skarbez argues that it is useful nevertheless to treat coherence as if it was objective, that is, that the programmed contents of the VR experience are geared towards matching user's expectations in a given context. Both Slater and Skarbez point out that PsI is not dependent on objective realism, but rather on what can be expected in a particular scenario \citep{skarbez2017survey,slater2022separate}.

In the VR literature, \textit{sense of embodiment} (SoE) is considered an illusion related to, but distinct from presence \citep{kilteni2012sense}. According to \cite{kilteni2012sense}, SoE consists of three components which can be manipulated using contemporary VR systems: a sense of body ownership (the body that I see belongs to me), a sense of agency (I am in control of this body), and a sense of self-location (a sense of one's location in relation to one's own body, not to be confused with PI). When a VR user is equipped with body-tracking and a virtual representation of a body viewed from a first-person perspective (1PP), a sense of body ownership towards the virtual body often emerges \citep{maselli2013building,kilteni2012sense}. Furthermore, a sense of body ownership towards physical stationary mannequin bodies have been elicited by synchronized visuo-tactile stimuli \citep{van2011being,van2016illusions}. Although SoE is considered an illusion of its own and not presence exactly, having a tracked 1PP virtual body appears to contribute towards presence as shown by decades of studies \citep{arora2022augmenting,slater2010simulating,skarbez2017psychophysical,slater1993representations,slater1995taking,slater1998influence}. According to some studies, a 1PP virtual body contributes towards both PI and PsI, specifically \citep{slater2009place, slater2010simulating,skarbez2017psychophysical}. In fact, SoE and presence are sometimes referenced in scientific literature interchangeably (e.g., \citealt{suzuki2023using}).

\subsection{Measuring presence}
Although presence is a phenomenon that can be easily understood once experienced, its empirical quantification has proven to be anything but straightforward. A common way in the past has been to utilize post-experimental questionnaires. These, however, pose multiple problems. Firstly, questionnaires force participants to reflect on their experience after the experience itself, when the actual illusion is already lost \citep{slater2022separate}. It has been argued that querying an abstract, and relatively uncommon (in day-to-day life) sensation such as presence in a virtual world in this way could even plant the sensation in the head of the participant even when it didn't exist in the first place \citep{slater2004colorful}. Furthermore, there are great differences in how questionnaires quantify presence; some questionnaires focus solely on PI as in \textit{sense of being there} \citep{slater1994depth,usoh2000using}, whereas others include related factors such as \textit{attention} and \textit{engagement} \citep{vasconcelos2016adaptation}, whereas some questionnaires focus not on merely the sensation itself, but include aspects of the VR experience that are assumed to affect presence (e.g., \citealt{witmer1998measuring}). Because of these discrepancies, comparing results between studies is difficult, and some questionnaires might not even be suitable for certain types of experiences like those that require relatively little interactivity \citep{skarbez2017survey,slater2022separate,kober2012using}. Questionnaires have also been criticized for their lack of sensitivity, especially when comparing systems of wildly different immersion levels; for example, researchers were not able to find a statistically significant difference in presence between a VR experience and physical reality in a between-group experience \citep{usoh2000using,slater2009place}. Although PI has been the focus of many presence questionnaires, there are no equally established questionnaires specifically targeting PsI. For Slater's PsI, past experiments have utilized study-specific questionnaires in which PsI is targeted by gauging whether participants physically or emotionally reacted to events of the VR experience as if they were real (e.g., \citealt{pan2016responses,neyret2020embodied}). Other examples targeting plausibility-related aspects of VR experiences include the \textit{reality judgment} questionnaire by Banos et al. and the questionnaire used by recent studies focusing specifically on the Latoschik \& Wienrich model of presence (e.g., \citealt{brubach2022breaking,brubach2024manipulating,banos2000presence}).

\textit{Breaks-in-presence} (BIPs) relate to occurrences in which the participant becomes more aware of the surrounding physical reality instead of the VR experience \citep{skarbez2017survey}. Self-reported BIPs have been used to model the presence of VR users probabilistically through discrete time-steps \citep{slater2000virtual} as well as a raw measure of presence (number of BIP occurrences correlating negatively with questionnaires; \citealt{brogni2003more}). Deliberate BIPs have also been utilized to experimentally manipulate presence for various research purposes, such as in order to investigate PI- and PsI-related aspects separately \citep{hofer2020role,brubach2022breaking,skarbez2020immersion,garau2008temporal}. Some studies have opted to utilize qualitative data instead of questionnaires. The studies of \cite{garau2008temporal} and \cite{poukePRESENCE} utilized thematic interviews to ask participants to describe their sense of presence and BIPs qualitatively, the latter also attempting to investigate PI and PsI separately. Furthermore, the studies of \cite{beacco2021disturbance} and \cite{slater2023sentiment} recruited participants to write about their experience in text and used sentiment analysis to investigate the results qualitatively. Finally, the so-called \textit{metamer} approach has been used to manipulate and model aspects of the VR experience (i.e., immersion and coherence factors) probabilistically to investigate their contribution to PI and PsI \citep{slater2010simulating,bergstrom2017plausibility,skarbez2017psychophysical}. 

\subsection{Behavioral and biometric measures}
Since presence is often understood as the participant responding to VR stimuli as if they were real \citep{sanchez2005presence,slater2009place}, it is often suggested that visible behavioral reactions (e.g., participant trying to dodge virtual objects), and physiological signals (e.g., changes in heart rate or skin conductance) could be used to measure presence \citep{skarbez2017survey}. For example, \cite{arora2022augmenting} found that when viewing a real-time 360° video experience through an HMD, participants equipped with a tracked virtual body (and who considered the said body realistic enough) reacted to stimuli approaching the 360° camera more strongly than without the body. Heart rate has been used as a proxy for experienced fear of heights in a VR experience \citep{meehan2002physiological} as well as a measure to identify BIPs \citep{garau2008temporal,skarbez2020immersion}. Skin conductance has been used to measure presence/body ownership  (the terminology for these illusions sometimes appear interchangeably within literature) when causing illusory threat to the virtual body  \citep{slater2009inducing,suzuki2023using}.

The problem with the measures above is that they require specific and intense stimuli in order to be effective; a VR user can experience strong presence even during mundane VR experiences that cause no measurable changes in visible behavior or stress-related physiological signals \citep{skarbez2017survey}. EEG offers a potential way to overcome these issues since it can also measure VR user's changes in cognition in real-time, while its sensitivity is not restricted to only overtly exciting or stress-inducing stimuli. Multiple studies have utilized the ERP technique with an auditory oddball paradigm based on the hypothesis that participants experiencing lower presence would allocate more attention toward the auditory stimuli compared to the VR experience (e.g., \citealt{kober2012using, grassini2021using, savalle2024towards}). Although the results have been promising, many scholars argue that presence cannot be equated with attention, as one can experience a strong sense of being in a place, yet not paying any particular attention to one's surroundings \citep{skarbez2017survey}. Based on the results of these studies, attention might serve as a useful proxy for presence similar to how, for example, heart rate can target presence indirectly through fearful stimuli but does not directly measure the sensation itself.  

\subsection{ERPs in Presence Research}
There are also ERP studies more directly targeting how contents of the experience itself cause changes in cognition. These studies commonly use some variation of BIPs and study their associated ERP signatures. The study of \cite{gehrke2019detecting} manipulated the timing congruency of feedback elicited when a participant was touching a virtual cube. They found that the amplitude of 
a negative going deflection around 200 ms after stimulus onset was modulated by the immersiveness of the feedback (visual only, haptic only, or visuo-haptic), which they dubbed the "prediction error negativity (PEN)". However, questionnaire results (IPQ; \citealt{vasconcelos2016adaptation}) indicated no significant differences in presence between the conditions \citep{gehrke2019detecting}. The study of \cite{padrao2016violating} manipulated body ownership by embodying participants an 1PP tracked virtual body that mimicked their movements in real-time in one condition and performed pre-recorded movements on some trials in another condition. In both conditions, participants were engaged in an error-prone fast-reaction task while the system sometimes gave erroneous feedback to participant's responses. They found an early frontocentral negativity to be sensitive to true self-generated errors and a late parietal negativity to be sensitive to non self-generated errors \citep{padrao2016violating}. The study of \cite{porssut2023eeg} focused on breaks in embodiment, specifically, and manipulated body ownership on the trial-level by causing unexpected hand movements in a virtual tracked body. They found the error-related negativity (ERN), Pe and N400 components to be sensitive to the manipulation \citep{porssut2023eeg}. The recent study of \cite{lehser2024feeling} also targeted BIPs caused by visuo-tactile mismatches. Participants observed a virtual hand corresponding to their real hand, and were instructed to press a button whenever a paintbrush touched their index finger. On a subset of trials, the paint brush was visually depicted to touch one finger while it touched another in the real world. They found a positive component at 120 ms to be sensitive to the mismatches, and similar to studies by \cite{porssut2023eeg} and \cite{padrao2016violating}, an N400-like component as well, which was interpreted as a visuo-tactile correlate of the N400 commonly observed in response to semantic violations.

A common limitation of the aforementioned studies is that none of the experiments made strong attempts to simulate a meaningful experience, but opted, instead, for utilizing visually minimal settings with little resemblance to commercial VR experiences, VR experiences commonly used in presence studies, or the real world. For example, \cite{gehrke2019detecting} speculate that their inability to detect an effect using questionnaire-based measures in their study could have been caused by the minimalistic experience failing to elicit presence in the first place.

In this paper, we present the results of our study investigating presence-related EEG signatures during a virtual arcade game experience. Our participants were given a task to achieve a high score in a slingshot game. They could perform the task embodied with a 1PP virtual body (embodied condition) or passively with game controllers and a static avatar (non-embodied condition). On most trials, the ball would fly from the sling obeying the regular physics of the game engine (normal trials), whereas on a minority of trials, we elicited SMC-related BIPs by freezing the ball and virtual slingshot (and, in the embodied condition, the avatar's hand) used to play the game (BIP trials). These manipulations resulted in a two-by-two within-subjects design. Based on previous findings and our own pilot data \citep{nardi2024quantifying}, we focused on three main components in our preregistration: the N2, the P3b, and the N400. 

\subsection{Theoretical Background of Relevant ERPs}

The family of early fronto-central negativities includes the mismatch negativity (MMN), associated with auditory sensory memory \citep{naatanen1978early, naatanen2001perception}, the error-related negativity (ERN/Ne), elicited in response to the noticing of self-generated errors \citep{falkenstein1990effects, gehring1993neural, gehring2007neural}, and the feedback-related negativity (FRN), elicited in response to performance error feedback \citep{miltner1997event, nieuwenhuis2004reinforcement}. There are a host of other N2 components associated with novelty \citep{courchesne1975stimulus}, perceptual mismatch \citep{kimura2006erp}, or error monitoring/cognitive control \citep{bartholow2005strategic}, and much debate over whether these and the aforementioned components reflect similar or different underlying processes \citep{folstein2008influence}. Early fronto-central negativities have been observed in response to SMC violations in some VR paradigms \citep{padrao2016violating, gehrke2019detecting}, but not others \citep{porssut2023eeg, lehser2024feeling}. If there is an electrophysiological signature of a mismatch between predicted and observed body movements, we would have good reason to expect it to manifest from the N2 family as a form of error-monitoring or perceptual mismatch. Further, the component should only be prevalent when violations occur in the presence of strong priors, such as controlling a body in a way that closely resembles our everyday experience. Such a component would thus make a strong candidate for an index of PI. 

Although canonically observed in the oddball paradigm and at times oversimplified as a sign of surprise \citep{donchin1981surprise}, the P300 has been demonstrated to arise in a variety of contexts beyond basic probability manipulations, supporting an interpretation that it reflects broader attentive processes. These processes have alternately been described as attentional resource allocation \citep{wickens1983performance}, context updating \citep{donchin1988p300}, information extraction \citep{gratton1990event}, and/or neural inhibition \citep{polich2007updating}, as well as evidence accumulation \citep{twomey2015classic}. A further distinction is made between attentional reorienting responses deriving from truly unexpected stimuli, as indexed by the P3a, and the integration of information deriving from behaviorally relevant stimuli, as indexed by the P3b \citep{polich2007updating}. In embodiment research, \cite{lehser2024feeling} observed a P300 in response to infrequent visuo-tactile mismatch events. Previously, \cite{gonzalez2014threat} reported a somewhat later but topographically consistent positivity in response to instances when a virtual knife passed through the participant's co-located virtual hand as opposed to passing through the table on which the hand rested (note that the events were equiprobable in this case). Despite the fact that a minority of the trials in our procedure contain BIP events, we expect that embodiment will be a stronger determinant of P3b amplitude than BIP probability, as relative to trials in the more passive non-embodied condition, trials in the embodied condition will offer richer behaviorally relevant information regarding body position that can contribute to the goal of achieving a high score \citep{begleiter1983p3}. As the P300 has been characterized as the process of noticing behaviorally relevant information and integrating that information into working memory (i.e., context updating; \citealt{donchin1988p300}), we question whether the component could represent a kind of bridge for information transfer between the representations maintaining a perceptually based PI and a cognitively based PsI, or whether it is better characterized as representing more general attentional processes that carry no specific relation to presence. 

The N400 is classically associated with semantic violations in language \citep{kutas1980reading, kutas2011thirty}. Extensions of this paradigm have found N400-like effects in response to semantically incongruent line drawings \citep{nigam1992n400, ganis1996search}, comic panels \citep{cohn2015getting}, and images \citep{ganis2003electrophysiological, vo2013differential}. \cite{tromp2018combined} demonstrated an N400 effect in virtual reality by creating mismatches between virtual objects in an environment and the spoken discourse regarding those objects by virtual agents. Meanwhile, others have reported N400-like components in response to embodiment manipulations \citep{padrao2016violating, porssut2023eeg, lehser2024feeling}, which have often been interpreted as violations of bodily semantics. We expect to see a similar type of N400 effect here, but we predict that the effect will extend beyond the body to violations of the semantics of physics laws (e.g., the freezing of a sling), appearing just as strongly to BIPs in the non-embodied condition as in the embodied condition. As the N400 has been described as reflecting semantic access \citep{kutas2011thirty}, a large N400 amplitude might be associated with breaks in PsI. 

In our experiment, we could not only replicate and extend multiple findings of previous studies within a single experiment, but we also did so by utilizing an immersive VR experience which, we argue, provides greater external validity for investigating a concept such as presence.
Finally, we discuss our findings in light of existing theories of presence, which we consider an interesting aspect overlooked by previous articles. 

\section{Hypotheses}
Confirmatory hypotheses and associated data collection and analysis protocols were preregistered at the Open Science Framework (\href{https://osf.io/g8f3e}{osf.io}). We predicted that:
\begin{itemize}
    \item HX: Embodiment questionnaire overall scores will be greater in the embodied condition than in the non-embodied condition.
    \item H1: The N2 component will be significantly more negative for trials with breaks in presence compared to normal trials only in the embodied condition but not in the non-embodied condition (interaction effect).
    \item H2: The P3b component will be significantly more positive in the embodied condition than the non-embodied condition, regardless of trial type (main effect). 
    \item H3: The N400 component will be significantly more negative for break in presence trials than normal trials, regardless of embodiment condition (main effect). 
\end{itemize}

This set of predictions thus sets up the idea that each highlighted component could reflect its own specific aspect of presence. Here, HX \footnote{We did not denote this hypothesis as H0 because that label is conventionally reserved for null hypotheses and here we predict an effect; however, as the effect is not particularly interesting on its own and serves mostly as a prerequisite for interpretation of other results, we have exceptionally denoted it as HX.} seeks to provide evidence that our embodiment manipulation increases the participant's sense of embodiment, using an established method. This evidence is necessary to ensure meaningful interpretations of the subsequent hypotheses. H1 then predicts that breaks in SMCs will be evident in the early N2 component, but only in the embodied condition where SMCs closely resemble real-life conditions, which would implicate the N2 as a functional marker of SMC breaks. H2 then predicts that the P3b will be larger in the embodied condition regardless of trial type, which could indicate that faithful embodiment leads to increased information extraction in the service of hitting targets. Finally, H3 predicts that the N400 will be larger for BIP trials independent of embodiment condition, suggesting a sensitivity to semantic violations that extends beyond language, images, and the body. 

\section{Methods}

\subsection{Participants}
Participants were recruited from the university and broader community using an online recruitment tool, which prevented participants who took part in pilot versions of the study from signing up. Effects sizes for H1:H3 were estimated in a pilot sample ($n = 12$) from a practically identical experiment to be very large (all Cohen's $dz > 1$) and thus \textit{a priori} power analysis was used to ensure greater than 95\% power to detect large ($dz >= 0.7$) effect sizes, resulting in a sample size of 30 participants (assuming within-subjects two-tailed t-tests with $\alpha = .05$). Our preregistered exclusion criteria stated that participants who failed to complete the whole experiment for any reason, such as their own withdrawal or irreparable equipment malfunction, and that participants for whom more than 40\% of epochs are flagged as containing artifacts, would be excluded and replaced prior to further analyses. Two participants were lost due to software malfunctions that interrupted their session and prevented them from completing the entire experiment, and one participant's data were found to be corrupted such that they could not be read for analysis. These three participants were replaced, resulting in a total of 33 participants of whom 30 passed exclusion criteria and were thus included in subsequent analyses. 

Of the 30 participants, 23 identified as male and seven identified as female when asked to self-identify their gender. All reported to be right-handed, to have normal or corrected-to-normal vision, and to have no history of traumatic brain injury. The sample had a mean age of 26.7 years (range: 18 to 37). When asked about their previous experience using VR technology, six reported to have never used it before, 13 reported to have used it only once or twice before, eight reported using it once or twice per year, and three reported using it once or twice per week. 

\subsection{VR Experience}
\begin{figure}[t]
\centering
\includegraphics[width=0.94\textwidth]{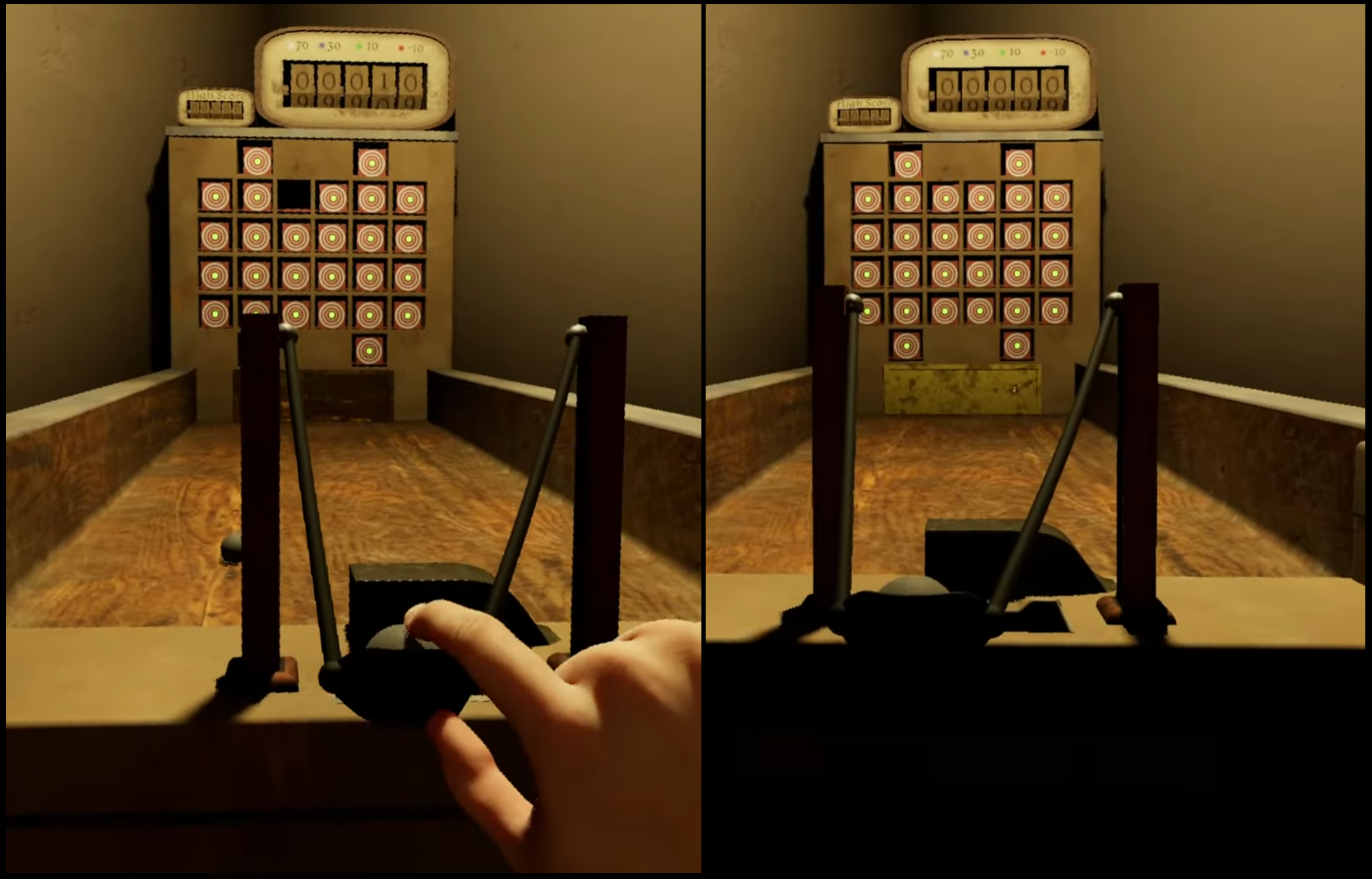}
\caption{The participants' task was to acquire as high a score as possible while playing a virtual arcade game. In the embodied condition (left), a virtual body was matched to participant's upper body movements and the participants could grab the slingshot naturally whereas in the non-embodied condition (right) the virtual body was stationary and the participants controlled the slingshot using game controllers.}
\label{Slingshot_both}
\end{figure}

The VR experience was developed in Unreal Engine 5.2 to act as our stimulus apparatus. The experience simulated a scenario in which participants were seated in front of an imaginary arcade game system in a furnished room. Playing this game system, the participants launched virtual balls with a slingshot while trying to hit targets placed in front of them (see Fig. \ref{Slingshot_both}). The targets were assigned different colors, indicating different scores upon being hit. The scores were assigned as -10 for red, +10 for green, +30 for blue, and +70 for white. The participants were instructed to try to get as high of a score as possible. In the beginning of each round, the colors of all targets were green, and a new color was assigned every time a ball struck the target. The frequencies for the colors were assigned as 30\% for red, 40\% for green, 20\% for blue, and 10\% for white. To prevent the game from becoming too challenging, red targets were reshuffled after 30 seconds. 

The experiment itself was divided into 10 rounds consisting of 200 trials, with the participant's score resetting between rounds. The first five trials of each condition (embodied or non-embodied) were considered practice trials. These trials were not included in analyses and the game board and score were reset after completion.

The participants viewed the VR experience with a Varjo Aero head-mounted display (HMD) and performed interactions using Valve Index controllers. In addition, the participants were wearing VIVE trackers strapped to their chests and elbows to establish full upper-body tracking together with the controllers and the HMD. Three Valve Index base stations were used for tracking the headset, controllers, and the trackers. We manipulated two key factors, embodiment and breaks-in-presence (BIPs), resulting in a two-by-two within-subjects design.

\subsubsection{Embodiment manipulation}
Throughout the experiment, the participants were given a first-person perspective (1PP) of a gender-matched virtual avatar developed with the Unreal Engine \textit{metahuman} tool. The participants were asked to match the avatar's skin tone as closely as possible to their own skin tone from five predefined options. Finally, the avatar's height was roughly matched to the participant's height from two alternatives (corresponding to ``average'' or ``tall''). Only the avatar's upper and lower bodies were modeled; their heads were not included in order to reduce the performance overhead of the application and to prevent visual artifacts in the field of vision. The virtual chair in which the participants were sitting was modeled to match the visual appearance, dimensions, and location of the physical chair in which they were sitting during the experiment. 

In the \textit{embodied} condition, the avatar's upper body tilt and hand and elbow transformations were matched to the participants' corresponding physical body movements. Furthermore, the avatar's hands and fingers were animated using a set of predefined skeletal animations responding to the participant pressing the trigger button of the controller, allowing the participant to grab the virtual ball and the slingshot from different angles. Besides indicating the grabbing visually, the motion controller also elicited a subtle vibration effect at the moment of grabbing. After grabbing the ball, the participant could then use his/her hand to pull the sling and adjust the ball's launch velocity (strength and direction), initiating the launch by releasing the trigger. The participants could, therefore, play the virtual game using a mechanism mimicking natural sensorimotor actions. Only right-handed participants were recruited and participants were instructed to use their right hand throughout the embodied condition, taking breaks as needed. Finally, the brass sign of the arcade game system was modeled to be slightly reflective to enhance body ownership \citep{gonzalez2010contribution}. The sign did not produce a perfect mirror reflection of the avatar body, but the participants could observe their movements to cause visible, synchronized reflections in the sign (See Fig. \ref{Slingshot_both} left).    

In the \textit{non-embodied} condition, we reduced the SMCs of the experience, aiming for reduced PI. In this condition, the virtual body was stationary; the upper body was leaning to the back of the chair, and its hands and arms were resting on the arms of the chair. Furthermore, instead of grabbing and controlling the slingshot with their hands, the participants used the thumbsticks of the Valve Index controllers to control the angle and the magnitude of the launch. The launch was initiated by pressing and releasing the trigger button of the Valve Index controller. The brass sign in this condition was non-reflective, so the participants could not even observe the reflections of the slingshot movement in the sign (See Fig. \ref{Slingshot_both} right). 

\subsubsection{BIP manipulation}
We manipulated SMCs in a randomly distributed 20\% of the trials, attempting to cause deliberate BIPs. These BIP-events were time-locked to the ball launch and lasted 1000 ms. In the \textit{embodied} condition, the Valve Index controller's transformation updates were frozen for the duration of the BIP event, causing the avatar's hand to temporarily freeze in place. Similarly, the slingshot was visually frozen in place with the ball remaining in its nest instead of exhibiting the usual animation coupled with the ball launch. The overall experience did not freeze, however, in terms of physics simulation updates; only the hand, the slingshot, and the ball were visually frozen in place. Instead, an invisible ball was launched according to the velocity specified at the time of the release, playing sounds and interacting with the rest of the game objects in an otherwise identical manner to the regular trials. After 1000 ms, the visibility of the balls was flipped: the active ball became visible, whereas the "prop" ball in the slingshot nest became invisible.

In the \textit{non-embodied} condition, the behavior of the BIP events was similar except for the freezing of the avatar's hands, since the avatar's hands were resting in place throughout the condition.

\subsection{EEG Recording}
EEG signals were measured using a 24 bit Brain Products actiCHamp amplifier (Brain Products GmbH, 2024), sampling continuous EEG data at 1000 Hz using active wet electrodes embedded in an elastic cap and placed according to the extended 10 percent system \citep{chatrian1985ten}. The ground electrode was placed at position Fpz. The actiChamp system delivers a ``reference free'' stream at the time of recording, with raw voltages recorded relative to the amplifier's internal ground. Impedances were kept below 10 $k\Omega$ before data was recorded inside an electrically shielded chamber. No filtering was applied during online recording. For synchronization of different data streams, the Lab Streaming Layer (LSL; \citealt{kothe2024lab}) was used, integrating event markers from Unreal Engine with the EEG data stream. Systems were connected via an offline local area network (LAN).

EEG data was processed using EEGLAB \citep{delorme2004eeglab} and ERPLAB \citep{lopez2014erplab}. Event markers were shifted 36 ms forward to account for photodiode-measured system latency, and data was downsampled to 250 Hz and rereferenced to the arithmetic average of the left and right mastoids. Ocular artifacts were removed via Independent Component Analysis (ICA; \citealt{makeig1997blind}). The data was first prepared for ICA by high-pass filtering using a fourth-order Butterworth filter with 12 db/oct rolloff with a 1.5 Hz half-amplitude cutoff, and data segments from channels (excluding frontal channels FP1, FP2, Fz, F3, F4, F7, and F8, so as to retain information specific to eye movements that would allow the ICA algorithm to accurately characterize them) exceeding +/-250 microvolts were removed using the continuous artifact detection function from ERPLAB with a 500 ms moving window and 250 ms step size, joining artifactual segments separated by less than 100 ms. ICA was performed on the prepared data using a compute unit device architecture (CUDA) implementation of Infomax ICA (CUDAICA; \citealt{raimondo2012cudaica}) on all channels excluding the mastoid reference channels. 

Subsequently, ICA weights were transferred to the unfiltered data and any components identified as deriving from ocular sources with more than 0.8 probability by IC label \citep{pion2019iclabel} were automatically removed. This ICA-cleaned data was then filtered using a fourth-order Butterworth filter with 12 db/oct rolloff with a 0.5 Hz half-amplitude cutoff. Because our primary measure of ERP components was a mean amplitude derived across prespecified latencies and electrode locations, no low-pass filter was applied \citep{zhang2024optimal}. 

ERPs were time-locked to the release of the slingshot, marking a ``trial''. The experiment was self-paced. The time required for a new ball to reset after a launch was a minimum of three seconds and it normally took participants several additional seconds to line up a subsequent shot. This allowed us to select long epoch windows. The data was binned into epochs beginning 1000 ms prior to event onset and lasting 1000 ms after event onset, using a baseline period from -1000 to -500 ms to help account for the response-locked (in our case, the release of a ball) nature of the time-locking event. 
Epochs containing voltage changes greater than $\pm$ 100 microvolts (after applying a 30 Hz low pass filter, used only for the purposes of this artifact detection step) were flagged using a 200 ms moving window with a 100 ms step size across the -1000 to 1000 ms interval. Clean epochs were averaged within conditions, within subjects. 

ERP regions of interest (ROIs) and latencies were selected based on pilot data (\citealt{nardi2024quantifying} and a subsequent unpublished pilot dataset of 12 participants which used an identical paradigm to the one reported here). N2 amplitude was quantified as the mean voltage from a fronto-central ROI composed of electrodes Fz, FC1, FC2, and Cz across the mean of the 220 to 300 ms latency window. P3b amplitude was quantified as the mean voltage from a central ROI composed of electrodes FC1, FC2, Cz, CP1, and CP2 across the mean of the 300 to 380 ms latency window. N400 amplitude was quantified as the mean voltage from the same central ROI composed of electrodes FC1, FC2, Cz, CP1, and CP2 across the mean of the 380 to 580 ms latency window.

\subsection{Procedure}
All procedures were in accordance with the Declaration of Helsinki and approved by the local ethical review board (ERB). Participants gave written informed consent prior to participating and were aware that they could withdraw at any time without consequence. Participants were given an oral overview of the complete procedure of the experiment, and then were shown to a restroom where they were asked to self-apply heart rate sensors according to a diagram on the wall (heart rate data was recorded for exploratory purposes and not processed in the current set of analyses). When the participant returned, they filled out a demographics questionnaire and we measured their head to fit an appropriately sized cap and start preparing the EEG recording system. Once impedances were brought to below 10 $k\Omega$, we presented participants with a live stream of their EEG data, highlighting their blinks, saccades, and jaw-clenching signals. Participants were instructed that they should behave normally other than to try to keep their face and neck muscles relaxed during the experiment. 

Participants were moved into the experimental room and the HMD was placed onto the participants head over the EEG cap. The HMD was modified with 3-D printed contact points which were padded with yoga mat cutouts in an effort to make the HMD more comfortable and compatible when used in conjunction with EEG electrodes. Participants were asked to adjust the HMD until the image was clear. Afterwards we performed another impedance check and returned any disturbed electrodes to below 10 $k\Omega$. SteamVR calibration was carried out for each participant separately to ensure their viewpoint was correctly registered to the virtual body. 

\subsection{Analyses}
Embodiment overall scores were calculated for each condition according to the method specified by \cite{peck2021avatar} and compared using a Wilcoxon signed-rank test (HX). Within-subjects ANOVAs (and pairwise follow-up tests, where appropriate) were used to test for the main effects and interactions for the breaks in presence and embodiment manipulations for each component (H1:H3). Alpha levels were set to .05.

Exploratory analyses included the extended Slater-Usoh Steed (SUS) questionnaire \citep{slater1994depth,pan2016responses} as a measure of PI, single-item post-experiment probes to gauge participants' conscious awareness of the BIP manipulation, as well as the application of mixed-effect models to determine whether questionnaire items or demographic information contribute to the successful prediction of ERP component amplitudes. 

For modeling purposes, the effect of BIPs was collapsed by subtracting mean BIP trial ERP responses from mean normal trials to account for the fact that questionnaire items were recorded at either the condition or experiment level rather than the trial level. Given the varied nature of predictor variables (some categorical, some numeric), we used linear mixed-effect models to predict this difference in ERP responses for various components. We used the `buildmer' package in R \citep{voeten2021package} for model selection. The algorithm was set to start with the maximal provided model, progressively remove terms, and compare competing models using likelihood ratio tests until a best-fitting model was found. Predictors included embodiment condition, embodiment score, SUS score, age, VR experience, condition preference, perceived difficulty, counterbalancing order, and responses to the aforementioned probes regarding participants' awareness of the BIP events. However, the question regarding items freezing was omitted due to probable misinterpretation, and gender was omitted due to only 7 of the 30 participants identifying as female. Interaction terms were prohibited due to the large number of predictors and limited number of instances at each level of interaction. A random intercept was provided to group responses within participants, and the algorithm was free to select random slopes and intercepts for all predictors. 

\section{Results}

\subsection{Confirmatory Results}
Embodiment overall scores were higher in the embodied condition ($Mdn = .73$) than in the non-embodied condition ($Mdn = .50$), $Z = 4.21$, $p = 2.76\times10^{-6}$, $r = .77$, providing support for the effectiveness of our embodiment manipulation (HX). 

\begin{figure}[!htbp]
\centering
\includegraphics[width=\textwidth]{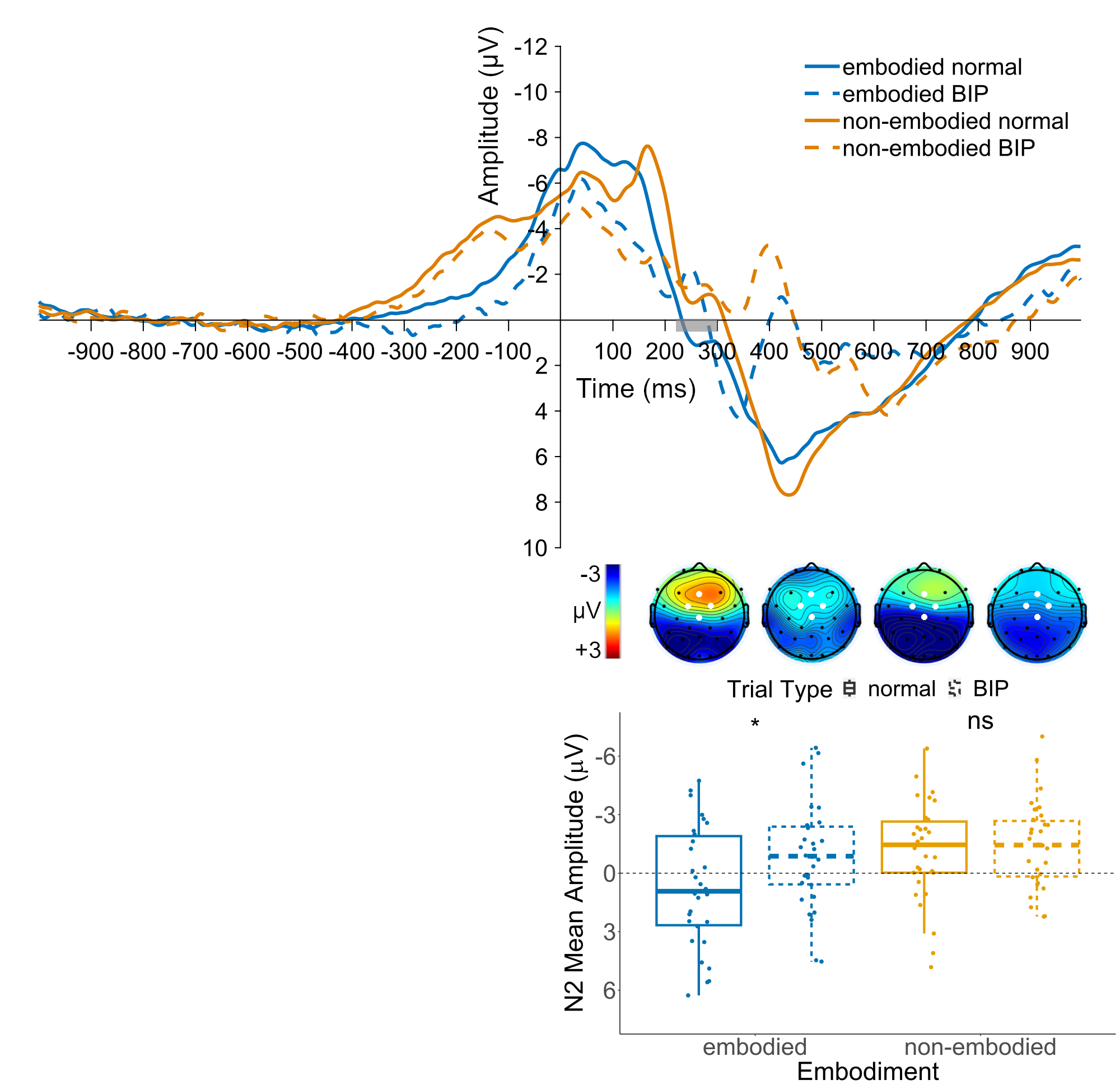}
\caption{Top: Grand average ERPs with embodied trials in blue, non-embodied trials in orange, normal trials in solid lines, and BIP trials in dashed lines. Time zero represents the moment that the ball is released. The latency window used in N2 component analyses is highlighted in gray across the x-axis (the same used to create the topographic and boxplots). Middle: Topographic plots displaying the mean voltage across the scalp for each condition within the latency window. Electrode locations forming the ROI for N2 component analyses (the same used to create the ERP and boxplots) are highlighted as white dots. Bottom: Boxplots depicting N2 distributions where dots represent individual subjects.} 
\label{figN2all}

\end{figure}

\begin{figure}[!htbp]
\centering
\includegraphics[width=\textwidth]{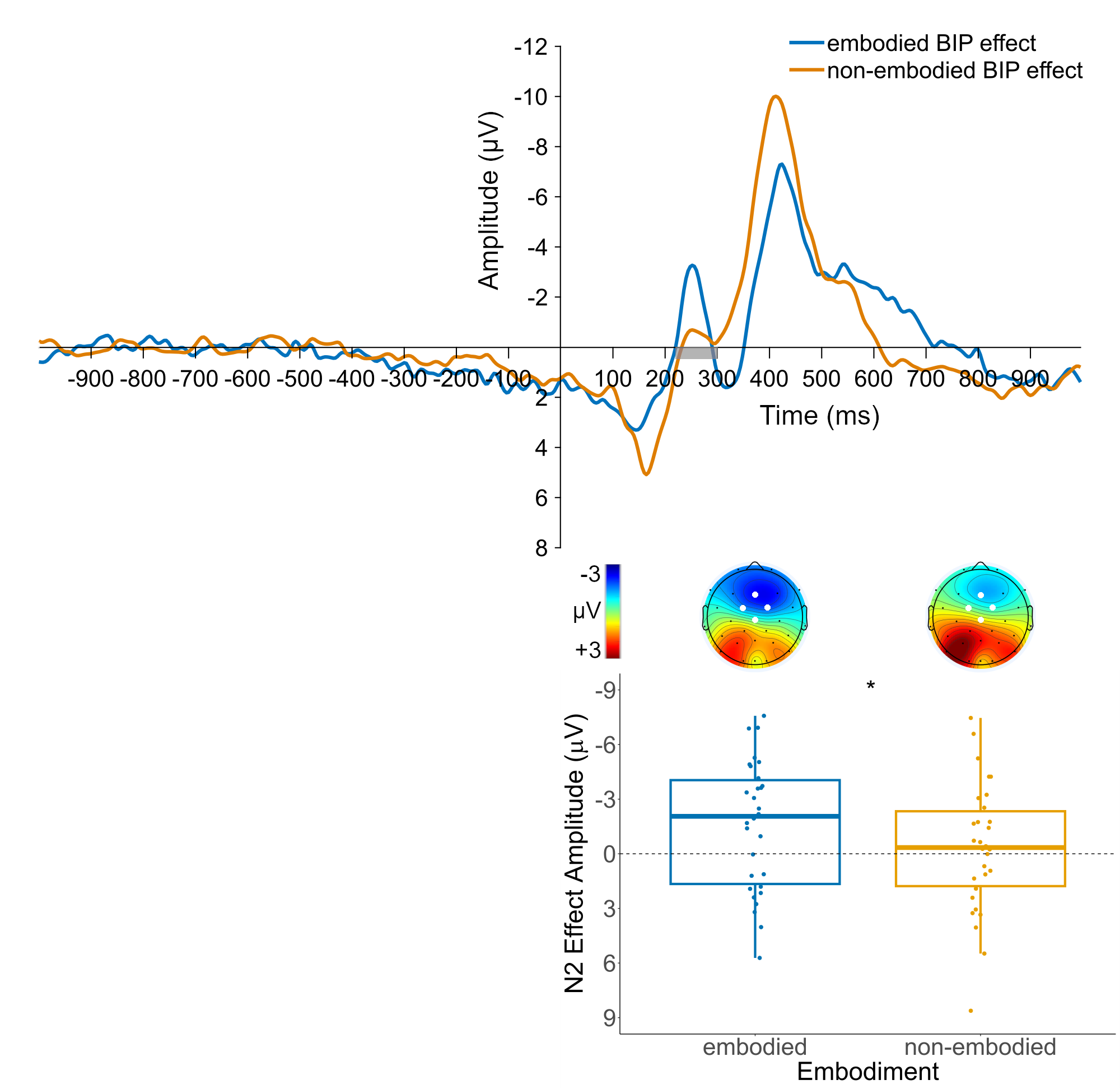}
\caption{Top: Difference waves (BIP trials minus normal trials) with embodied trials in blue and non-embodied trials in orange. Time zero represents the moment that the ball is released. The latency window used in N2 component analyses is highlighted in gray across the x-axis (the same used to create the topographic and boxplots). Middle: Topographic difference plots displaying the mean voltage across the scalp for each condition effect within the latency window. Electrode locations forming the ROI for N2 component analyses (the same used to create the ERP and boxplots) are highlighted as white dots. Bottom: Boxplots depicting N2 effect distributions where dots represent individual subjects.} 
\label{figN2diff}
\end{figure}

N2 mean amplitudes indicated a main effect of embodiment, $F(1,29) = 7.69$, $p = .010$, $\eta_{p}^{2} = .21$, and a significant trial type by embodiment interaction, $F(1,29) = 5.42$, $p = .027$, $\eta_{p}^{2} = .16$, but no main effect of trial type, $F(1,29) = 2.57$, $p = .12$, $\eta_{p}^{2} = .08$ (Figure \ref{figN2all}). Post-hoc tests revealed a more negative N2 in response to the BIP trials than to normal trials in the embodied condition, $m_{diff} = -1.57$, $t(29) = -2.43$, $p = .021$, $g = -0.43$, which was not apparent in the non-embodied condition, $m_{diff} = -0.31$, $t(29) = -0.47$, $p = .64$, $g = -0.08$ (Figure \ref{figN2diff}), thus supporting our hypothesis that the N2 response to a BIP largely depends on embodiment (H1).

\begin{figure}[!htbp]
\centering
\includegraphics[width=\textwidth]{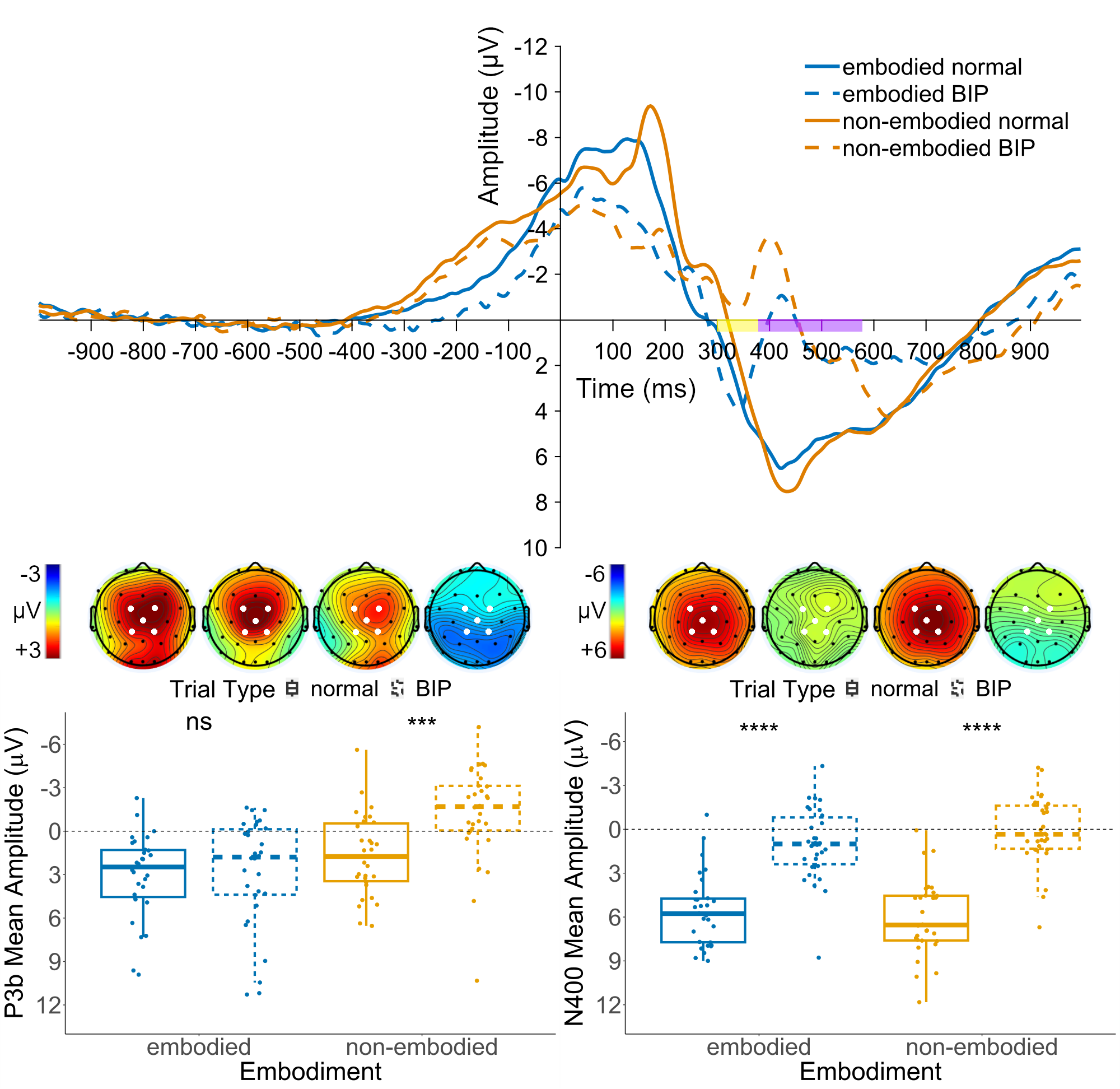}
\caption{Top: Grand average ERPs with embodied trials in blue, non-embodied trials in orange, normal trials in solid lines, and BIP trials in dashed lines. Time zero represents the moment that the ball is released. The latency window used in component analyses is highlighted across x-axis in yellow for the P3b and purple for the N400 (the same as those used to create the topographic and boxplots). Middle: Topographic plots displaying the mean voltage across the scalp for each condition within the latency window for the P3b (left) and N400 (right). Electrode locations forming the ROI for P3b and N400 component analyses (the same used to create the ERP and boxplots) are highlighted as white dots. Bottom: Boxplots depicting P3b (left) and N400 (right) distributions where dots represent individual subjects.} 
\label{figP3N4all}
\end{figure}

\begin{figure}[!htbp]
\centering
\includegraphics[width=\textwidth]{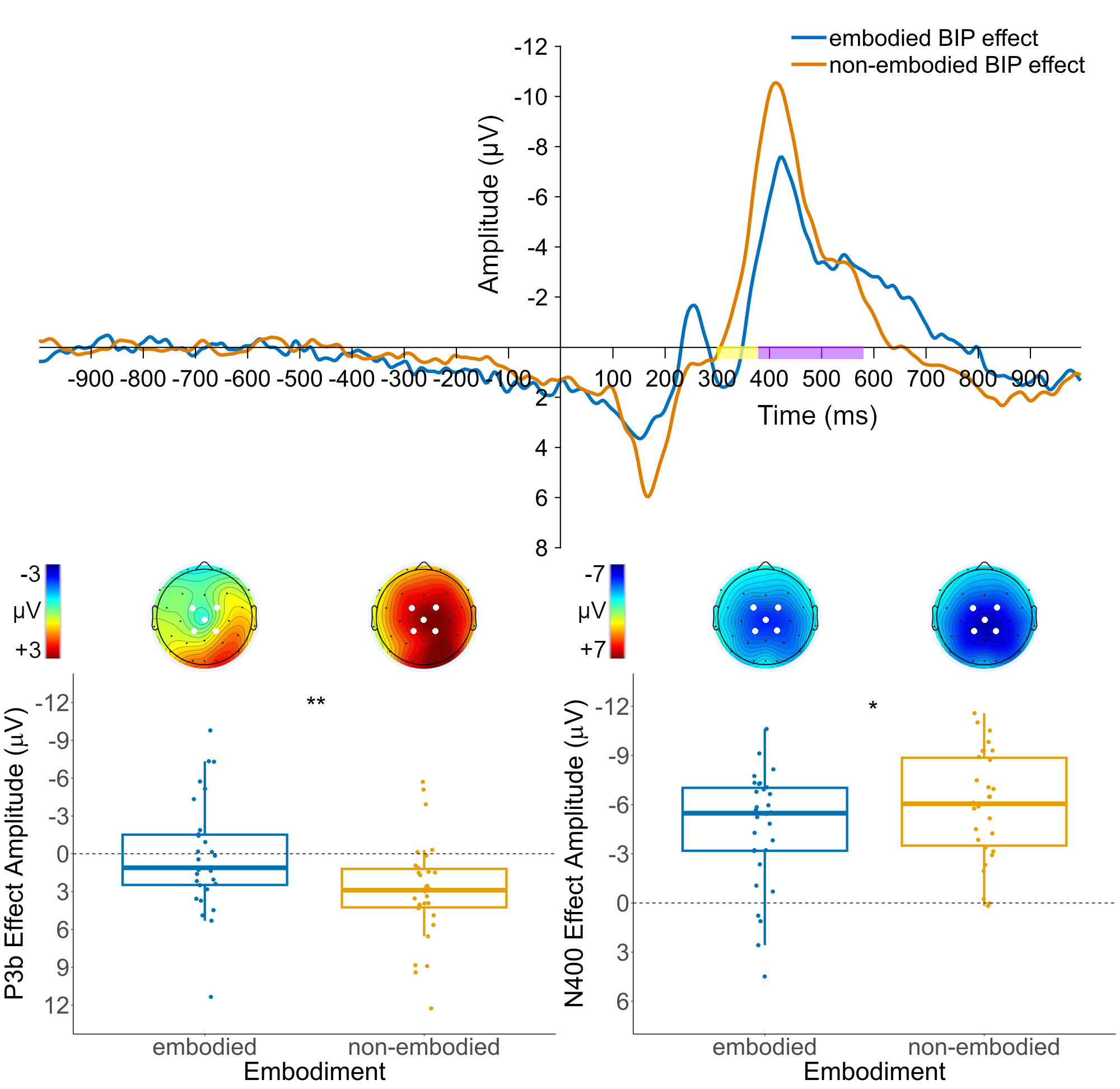}
\caption{Top: Difference waves (BIP trials minus normal trials) with embodied trials in blue and non-embodied trials in orange. Time zero represents the moment that the ball is released. The latency window used in component analyses is highlighted across x-axis in yellow for the P3b and purple for the N400 (the same as those used to create the topographic and boxplots). Middle: Topographic difference plots displaying the mean voltage across the scalp for each condition effect within the latency window for the P3b (left;  exceptionally depicted as normal trials minus BIP trials) and N400 (right). Electrode locations forming the ROI for P3b and N400 component analyses (the same used to create the ERP and boxplots) are highlighted as white dots. Bottom: Boxplots depicting P3b (left; exceptionally depicted as normal trials minus BIP trials) and N400 (right) effect distributions where dots represent individual subjects.} 
\label{figP3N4diff}
\end{figure}

P3b mean amplitudes showed a significant main effect of trial type, $F(1,29) = 5.68$, $p = .024$, $\eta_{p}^{2} = .16$, a significant main effect of embodiment, $F(1,29) = 25.22$, $p = 2.38\times10^{-5}$, $\eta_{p}^{2} = .47$, and a significant trial type by embodiment interaction, $F(1,29) = 12.44$, $p = .001$, $\eta_{p}^{2} = .30$ (Figure \ref{figP3N4all}). Post-hoc tests revealed that P3b amplitudes were significantly larger in the embodied than non-embodied conditions for both BIP trials, $m_{diff} = 4.07$, $t(29) = 5.06$, $p = 2.14\times10^{-5}$, $g = 0.90$, and normal trials, $m_{diff} = 1.36$, $t(29) = 2.81$, $p = .009$, $g = 0.50$. Furthermore, there was in general a \textit{smaller} P3b in response to BIP trials compared to normal trials, however this effect was only significant in the non-embodied condition, $m_{diff} = -2.93$, $t(29) = -4.02$, $p = 3.76\times10^{-4}$, $g = -0.72$, and not in the embodied condition, $m_{diff} = -0.22$, $t(29) = -0.27$, $p = .78$, $g = -0.05$ (Figure \ref{figP3N4diff}). Thus, while the P3b response was primarily driven by embodiment (H2), a particular lack of a response for BIP trials in the non-embodied condition produced a significant interaction (see Figure \ref{figP3N4all}), giving us only partial support for H2.

For N400 mean amplitudes, we observed a significant main effect of trial type, $F(1,29) = 91.22$, $p = 1.85\times10^{-10}$, $\eta_{p}^{2} = .76$, and a significant trial type by embodiment interaction, $F(1,29) = 5.43$, $p = .027$, $\eta_{p}^{2} = .16$, but no main effect of embodiment, $F(1,29) = 0.71$, $p = .41$, $\eta_{p}^{2} = .02$ (Figure \ref{figP3N4all}). Post-hoc tests indicated more negative N400 responses to the BIP trials than normal trials in both the embodied condition, $m_{diff} = -4.54$, $t(29) = -7.03$, $p = 9.78\times10^{-8}$, $g = -1.25$, and the non-embodied condition, $m_{diff} = -5.94$, $t(29) = -9.84$, $p = 9.51\times10^{-11}$, $g = -1.75$ (Figure \ref{figP3N4diff}). Further contrasts revealed that BIP trials produced slightly more negative responses in the non-embodied condition than in the embodied condition, $m_{diff} = -0.92$, $t(29) = -2.06$, $p = .048$, $g = -0.37$, whereas this difference was not observed for normal trials,  $m_{diff} = 0.48$, $t(29) = 1.41$, $p = .17$, $g = 0.25$. Similarly to our P3b effects, we find that this pattern of results supports our hypothesis that the N400 is overwhelmingly driven by BIP events (H3), while conceding that embodiment appears to make a modest yet statistically significant contribution. 

\subsection{Exploratory Findings}
Though they were not planned components of interest in our preregistration, there were two visually striking earlier components that can be observed in both Figures \ref{figN2all} and \ref{figP3N4all} which are thus worth mentioning. Leading up to the release of the ball, one can see a slow-going negative buildup that is maximal over central electrode sites indicative of a readiness potential, which is visually larger for the non-embodied condition. Such potentials are common in situations when participants are preparing a response or anticipating a stimulus (see \citealt{brunia2011negative}). Then, prior to the N2 window, an early negative potential appeared between 120 to 220 ms that was maximal over parietal and occipital electrode sites and prominent only in normal trials. We interpret this component as a visually-evoked potential (VEP) attributable to the visual difference between what the participant sees when a trial is frozen versus when the trial proceeds normally with motion of the ball, sling, and hand. 

SUS mean scores were significantly higher in the embodied condition ($Mdn = 5.33$) than in the non-embodied condition ($Mdn = 4.58$), $Z = 3.47$, $p = 2.46\times10^{-4}$, $r = .63$, indicating that operating an embodied avatar boosted participants' sense of place illusion. When asked about their preference on interaction methods, 24 participants preferred playing the game while controlling the virtual body, while 6 preferred playing using just the thumbsticks. For those who preferred the controlling the virtual body, many open responses cited greater control or engagement, whereas those who preferred thumbstick control tended to cite greater fatigue controlling the virtual body as their main motivator. There was no significant difference in perceived difficulty between the embodied ($Mdn = 4$) and non-embodied ($Mdn = 4.5$) conditions, $Z = -0.83$, $p = .42$, $r = .15$. 

We were curious to what degree participants would consciously register the BIP events. When asked whether there were any moments when participants became suddenly aware of the real-world laboratory, eight affirmed this never happened, nine only without the virtual body, two only with the virtual body, and 11 affirmed it happened in both cases. The 15 provided open responses to what triggered these moments varied greatly in content, however the modal response (from four participants) concerned the discomfort of the equipment and/or fatigue. When asked whether they noticed any bugs or errors, two claimed to have not noticed any, zero only without the virtual body, six only with the virtual body, and 22 affirmed to have noticed in both cases. Twenty-five of the 30 accompanying free responses described bugs that could plausibly relate to the BIP manipulation. Finally, when explicitly asked whether they noticed any instances of items freezing, 13 (!) claimed to have not noticed any such instances, one only without the virtual body, four only with the virtual body, and 12 affirmed to have noticed such instances in both cases. Given the pattern of open responses to the previous question, we suspect that the large number of negative responses to this item might stem from a misunderstanding of the question phrasing as none of the participants were native English speakers. 

\begin{figure}[!htbp]
\centering
\includegraphics[width=100mm]{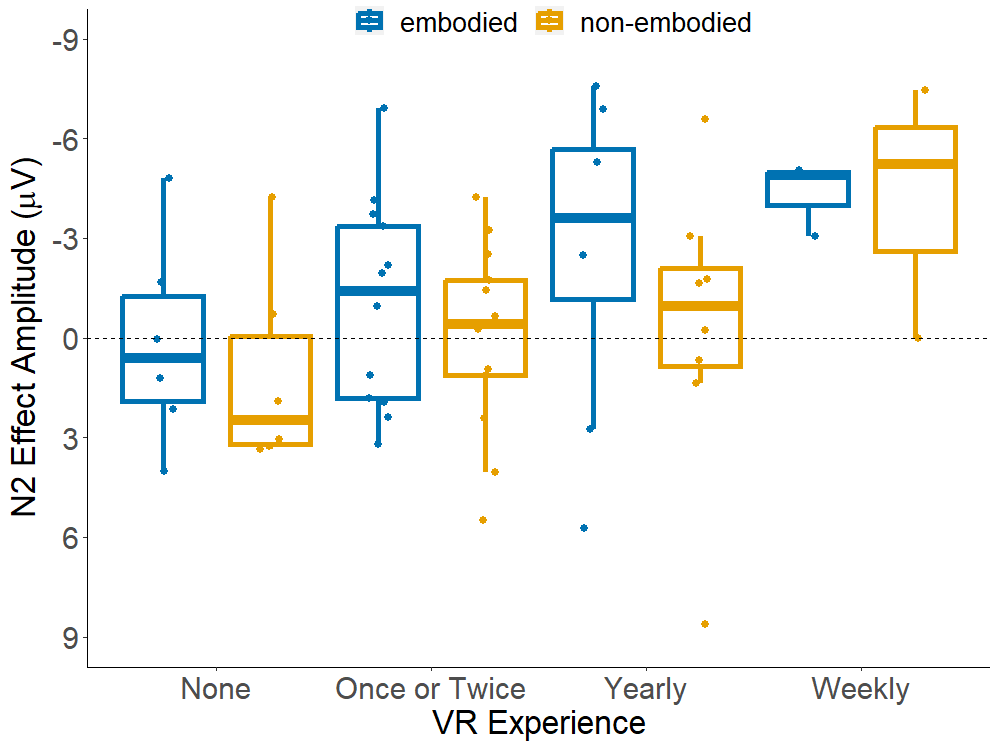}
\caption{Boxplots depicting N2 effect distributions according to VR experience, where dots represent individual subjects. Embodied responses are depicted in blue whereas non-embodied responses are depicted in orange.}
\label{figN2byVR}
\end{figure}

For the P3b and N400 responses, the best-fitting model was simply the embodiment condition predictor with a random intercept for participants, similar to the analyses already presented in the confirmatory results section. However, for the N2 response, the best fitting model included VR experience ($\beta = -1.18$; $95\%$ $CI = [-2.14, -.23]$), embodiment ($\beta = -1.27$; $95\%$ $CI = [-2.35, -.18]$) and a random intercept for participants ($\sigma^2 = 6.62$). This model was a significantly better fit in comparison to the null model, $\chi^2(2,N=30) = 10.84$, $p = .004$, and the embodiment condition-only model, $\chi^2(1, N=30) = 5.70$, $p = .017$. The slope of the VR experience predictor would indicate that the N2 effect increases with increasing experience using VR systems. This trend can be clearly observed in Figure \ref{figN2byVR} which plots N2 effects by VR experience.

\section{Discussion}
\subsection{Hypotheses}
Our results mostly supported our hypotheses HX-H3. Support for HX validated that our body ownership manipulation was successful, and that the tracked virtual body elicited a larger body ownership than the non-tracked one (embodied vs non-embodied conditions). This allows us to interpret the main results meaningfully. Our SMC manipulation did not only depend on the visual representation of body movements, but also on how the participants controlled the slingshot. Our exploratory SUS results also indicate that the participants experienced greater PI during the embodied condition, specifically.

Regarding H1, our results showed a larger BIP N2 response during the embodied condition than the non-embodied one, supporting this hypothesis. A similar finding was also made by \cite{gehrke2022neural} in which N2 BIP-effect magnitude was dependent on the fidelity of sensory feedback during a VR motor task. Furthermore, a BIP-like effect within the N2 region was also found in studies by \cite{padrao2016violating}, \cite{lehser2024feeling}, and \cite{porssut2023eeg}.  Given the theoretical implication that PI would depend on the SMCs supported by the system, these findings support the idea that N2 could serve as an index for breaks in PI.

We also partially supported H2, according to which the P3b component was significantly more positive in the embodied condition compared to the non-embodied condition. Contrary to our predictions, however, there was also a main effect of trial type and a small interaction effect according to which P3b amplitude was smallest for BIP trials in the non-embodied condition. In neuroscientific literature, the P3b has been associated with information extraction and attention. Previous presence EEG studies have utilized P3b as a proxy for presence through attention using auditory oddball tasks (e.g., \citealt{savalle2024towards}); even if VR literature cautions against coupling attention directly with presence (\citealt{skarbez2017survey}). In these paradigms, the increased P3b effect for oddballs indicates the shift of attention towards an external stimulus. In this experiment, however, we did not utilize a classical oddball task or attempt to explicitly manipulate attention. Though there was a frequency manipulation in terms of BIP trials occupying only 20\% of all trials, our P3b was either not sensitive at all to less frequent trials, or elicited a smaller amplitude in these trials, which is not what is observed in typical P3b oddball designs. In fact, the largest P3b effect size by a considerable margin was at the block-level in which the P3b magnitude was greater for embodied trials. This overall pattern is not difficult to interpret if viewed from the standpoint of information extraction. The participant's goal in our task was only to achieve a high score, for which they needed to focus on their manipulation of the sling and then take into account the immediate consequences of that manipulation regarding whether it led to a hit or a miss. The manual manipulation required by the embodied condition offered more sensitive strategies and therefore more information to encode than the relatively coarse manipulation offered by the thumbsticks in the non-embodied condition. The BIP trials then cut that information extraction process short with a nonsensical event that interrupts the player's progress. It could have been the case that the freezes (BIP events) in the embodied condition were interpreted as something the players did themselves, prompting them to put more effort (consciously or unconsciously) into encoding the preceding events, whereas the freezes in the non-embodied condition could have been more easily externalized as a glitch in the game that was independent of the player, and thus not worthy of encoding. 

The matter is somewhat complicated by what appears to be a delayed P3b in normal trials, peaking near 400 ms, in contrast to the rapidly reversing P3b in BIP trials that seems give way for the N400 component. We should note that the early portion of this delayed P3b is captured within our P3b analysis window while the rest of the activity continues on into the next window, likely exaggerating the magnitude of the subsequent N400 effect to some degree. Accounting for differences in peak timing, the P3b peak was largest for normal trials, whether embodied or non-embodied, smaller for embodied BIP trials, and almost nonexistant for non-embodied BIP trials. According to visual inspection (comparing the normal trials in the N400 portions of Figure \ref{figP3N4all} to the BIP trials of the P3b portions of Figure \ref{figP3N4all}), the delayed timing of the normal trials thus maintains the overall pattern from the original analysis window but shifts the main effect emphasis from embodiment to BIPs, while seemingly preserving the interaction. The equivalence of the normal trials in this later window across levels of embodiment, which, recall, did differ in terms of SUS scores, suggests that the P3b is not specifically sensitive to some aspect of presence \textit{per se}. We instead contend that a general information extraction account is still the most parsimonious explanation; increased immersion (i.e., richer sensorimotor contingencies) and coherence (i.e., the slingshot working as it should) conferred a greater amount of task-relevant info to the participant, and therefore, a larger P3b was generated.

Regarding H3, we found, as predicted, that the N400 component was significantly more negative for the BIP trials. However, we also found a small but significant interaction effect, according to which the N400 BIP effect was greater for the non-embodied trials. This result did not precisely match our predictions, according to which N400 would not have been sensitive to body ownership. It is also surprising that the N400 effect was greater for the non-embodied condition, specifically. As the N400 is commonly associated with language incongruencies and disruptions of visual semantics, it is tempting to associate this effect with coherence and PsI - perhaps the freezing of the slingshot and the disappearing ball elicited the N400 effect? However, since previous VR studies have also found a N400 effect when disrupting tracked full-body representations \citep{padrao2016violating,porssut2023eeg,lehser2024feeling}, we cannot be certain whether this wasn't the case in this study as well. Although the N400 was elicited (and even larger in magnitude) in the non-embodied condition, our experimental manipulation did not include a condition in which the slingshot would have frozen while the hand motion was retained as usual.

A second potential interpretation that N400 serves as an index for body movement violations (\citealt{padrao2016violating,porssut2023eeg,lehser2024feeling}; but is not, perhaps, sensitive to touch-related sensory feedback violations utilized by \citealt{gehrke2022neural}), is not entirely supported by our results either because of the non-embodied N400 effect stated above. This leads into a third interpretation: perhaps playing the slingshot game with the thumbsticks was more predictable compared to natural sensorimotor actions. This could have led to stronger predictions and a larger prediction error at BIP events. A fourth interpretation is that breaking SMCs sends a cascade of prediction errors across various levels of cognition even without coherence-specific manipulations. In terms of presence theory, this would imply that a break in PI would always result in a break in PsI as well as suggested by some behavioral studies \citep{poukePRESENCE,brubach2024manipulating}. The missing N400 effect in the study of \cite{gehrke2022neural} could therefore result from the lack of a meaningful environment and a task, instead of a missing body representation. It could be that the N400 effect observed in this study is a mix of the above, and further investigation is necessary to disentangle the exact causes for the effect found in this study.

\subsection{Exploratory Findings and Limitations}
As stated above, we found that SUS scores were significantly higher during the embodied condition. This is in line with multiple previous studies according to which a visual representation of a body and the utilization of natural sensorimotor actions in VR increase presence (e.g., \citealt{slater1995taking,slater1993representations,arora2022augmenting}). Furthermore, most participants reported the embodied condition being more enjoyable while there was no significant difference in perceived difficulty. Interestingly, our exploratory analysis found that the N2 effect was larger for participants with more experience in VR. It could be that experience with VR controllers and hardware allowed participants to immerse themselves into the experience without consciously thinking about the controllers, or similarly that more experience in virtual environments allowed predictive coding processes to fine-tune expectations regarding SMCs in VR, leading to larger responses when they were violated. On the other hand, this finding suggests that the capability of VR to elicit PI would not diminish through increased exposure to VR. 

We also queried whether the participants consciously registered our BIP manipulations, and whether they caused phenomenological losses in presence. Although 25 participants reported bugs that appear to relate to our BIP manipulations, most of the participants attributed losses in the \textit{"sense of being there"} to fatigue or discomfort related to the EEG and VR equipment. Interestingly, 11 participants reported not noticing any instances of items freezing in place, which contradicts their other reports related to bugs and glitches. We suspect that this discrepancy could be related to language. Based on the responses above, although our EEG results are quite robust and sensitive to our manipulations, we cannot state with perfect certainty whether our N2 and N400 effects are strictly associated with conscious losses in presence. Instead, these effects could also index the perception of glitches in the VR experience which the brain processes either unconsciously, or which the VR user immediately forgets. We should also note that the relative lack of women in our sample limits the generalizability of our findings and should better balanced in future experiments.

Finally, this study is limited by the fact that our experimental paradigm was not designed to cause distinct breaks in PI and breaks in PsI, specifically, and we, therefore, cannot fully isolate the neural signatures of either. Although it is tempting, based on existing literature, to infer that the N2 response is specific to breaks in PI, and that the N400 response is specific to breaks in PsI, a manipulation that generates and breaks each illusion independently would be required to draw this conclusion with certainty. 
However, empirical separation of PI and PsI appears to be a nontrivial task \citep{slater2022separate}, and it is occasionally disputed whether PsI can even exist without PI in a VR experience (e.g. \citealt{poukePRESENCE, brubach2024manipulating}). A potential pathway for stronger presence manipulation would be by comparing experiences utilizing traditional displays to ones using HMDs (i.e. non-immersive and immersive displays, respectively), however this is difficult to achieve in the ERP paradigm where one typically wants to minimize visual differences between conditions. That said, future studies should look for novel ways to target either PI-related or PsI-related aspects of the virtual experience to advance the separation of their neural signatures.

\section{Conclusion}
In this paper, we reported the results of our preregistered ERP experiment focusing on neural signatures indicating either existence or loss of presence in VR. Participants were engaged in a virtual arcade game in which their task was to get as high a score as possible shooting balls towards targets using a slingshot while EEG data was collected during the gameplay. 

We caused deliberate BIPs at 20\% of the trials by freezing the slingshot and the participant's hand in the embodied condition, and the slingshot only in the non-embodied one. 

We found several ERP components to be sensitive to our manipulations, partially confirming our preregistered hypotheses. We found that N2 component was sensitive to BIP trials, and also that the effect was significantly stronger during the embodied condition. We also found that P3b was more positive for all trial types during the embodied condition. However, we also found that the P3b was less positive for the BIP trials although this effect was significant only within the non-embodied cognition. We also found that the N400 component was sensitive to BIP trials, however, although we predicted that the N400 would not be sensitive to embodiment, we found a small effect according to which the N400 was slightly more negative during the non-embodied condition. 
We suggest that the N2 component could serve as an index for breaks in SMCs, and therefore PI. We also suggest that P3b does not relate specifically to presence, but rather serves as separate but related index of, in our case, information extraction, or in other paradigms, attentional reorienting. We also replicated findings from previous studies according to which N400 is sensitive to virtual body visuomotor violations, however, here we show that the N400 also extends to violations of physics that are independent of bodily sensations. There are multiple interpretations for our N400-related findings. We suggest that future research focuses on decoupling the role of the N400 component in relation to body movement violations and the coherence of the virtual environment.   

Despite acknowledged limitations, this work advances existing ERP presence research in multiple ways. In our experimental paradigm, we manipulated presence independently through embodiment, allowing a cause-and-effect investigation of presence-related ERPs. We were also able to validate the effectiveness of our manipulation through traditional questionnaires. Our trial level manipulation targeted sensorimotor contingencies, indexing presence through ERPs coupled with BIPs instead of measuring presence indirectly through attention. We also utilized a realistic, aesthetically pleasing virtual environment with a meaningful task to minimize confounds due to the experiment itself failing to elicit presence. Finally, we reflected our findings against theories of presence, which has been uncommon in previous neuroscientific presence investigations.

\section{Acknowledgements}
We would like to thank Alexis Chambers, Max Garcia, Aleksi Heikkilä, Ata Jodeiri Seyedian, Aleksi Sierilä, and Hung Trinh for their dedicated data collection efforts. 

Funding: This work was supported by the European Research Council (project ILLUSIVE 101020977), the Research Council of Finland (projects BANG! 363637, and PERCEPT 322637), and the University of Oulu and Research Council of Finland (PROFI7 352788).

\bibliographystyle{elsarticle-harv} 
\bibliography{cas-refs}

\begin{thebibliography}{96}
\expandafter\ifx\csname natexlab\endcsname\relax\def\natexlab#1{#1}\fi
\providecommand{\url}[1]{\texttt{#1}}
\providecommand{\href}[2]{#2}
\providecommand{\path}[1]{#1}
\providecommand{\DOIprefix}{doi:}
\providecommand{\ArXivprefix}{arXiv:}
\providecommand{\URLprefix}{URL: }
\providecommand{\Pubmedprefix}{pmid:}
\providecommand{\doi}[1]{\href{http://dx.doi.org/#1}{\path{#1}}}
\providecommand{\Pubmed}[1]{\href{pmid:#1}{\path{#1}}}
\providecommand{\bibinfo}[2]{#2}
\ifx\xfnm\relax \def\xfnm[#1]{\unskip,\space#1}\fi
\bibitem[{Arora et~al.(2022)Arora, Suomalainen, Pouke, Center, Mimnaugh, Chambers, Pouke and Lavalle}]{arora2022augmenting}
\bibinfo{author}{Arora, N.}, \bibinfo{author}{Suomalainen, M.}, \bibinfo{author}{Pouke, M.}, \bibinfo{author}{Center, E.G.}, \bibinfo{author}{Mimnaugh, K.J.}, \bibinfo{author}{Chambers, A.P.}, \bibinfo{author}{Pouke, S.}, \bibinfo{author}{Lavalle, S.M.}, \bibinfo{year}{2022}.
\newblock \bibinfo{title}{Augmenting immersive telepresence experience with a virtual body}.
\newblock \bibinfo{journal}{IEEE Transactions on Visualization and Computer Graphics} \bibinfo{volume}{28}, \bibinfo{pages}{2135--2145}.
\bibitem[{Ba{\~n}os et~al.(2000)Ba{\~n}os, Botella, Garcia-Palacios, Villa, Perpi{\~n}{\'a} and Alcaniz}]{banos2000presence}
\bibinfo{author}{Ba{\~n}os, R.M.}, \bibinfo{author}{Botella, C.}, \bibinfo{author}{Garcia-Palacios, A.}, \bibinfo{author}{Villa, H.}, \bibinfo{author}{Perpi{\~n}{\'a}, C.}, \bibinfo{author}{Alcaniz, M.}, \bibinfo{year}{2000}.
\newblock \bibinfo{title}{Presence and reality judgment in virtual environments: a unitary construct?}
\newblock \bibinfo{journal}{CyberPsychology \& Behavior} \bibinfo{volume}{3}, \bibinfo{pages}{327--335}.
\bibitem[{Bartholow et~al.(2005)Bartholow, Pearson, Dickter, Sher, Fabiani and Gratton}]{bartholow2005strategic}
\bibinfo{author}{Bartholow, B.D.}, \bibinfo{author}{Pearson, M.A.}, \bibinfo{author}{Dickter, C.L.}, \bibinfo{author}{Sher, K.J.}, \bibinfo{author}{Fabiani, M.}, \bibinfo{author}{Gratton, G.}, \bibinfo{year}{2005}.
\newblock \bibinfo{title}{Strategic control and medial frontal negativity: Beyond errors and response conflict}.
\newblock \bibinfo{journal}{Psychophysiology} \bibinfo{volume}{42}, \bibinfo{pages}{33--42}.
\bibitem[{Beacco et~al.(2021)Beacco, Oliva, Cabreira, Gallego and Slater}]{beacco2021disturbance}
\bibinfo{author}{Beacco, A.}, \bibinfo{author}{Oliva, R.}, \bibinfo{author}{Cabreira, C.}, \bibinfo{author}{Gallego, J.}, \bibinfo{author}{Slater, M.}, \bibinfo{year}{2021}.
\newblock \bibinfo{title}{Disturbance and plausibility in a virtual rock concert: A pilot study}, in: \bibinfo{booktitle}{IEEE VR}, \bibinfo{organization}{IEEE}. pp. \bibinfo{pages}{538--545}.
\bibitem[{Begleiter et~al.(1983)Begleiter, Porjesz, Chou and Aunon}]{begleiter1983p3}
\bibinfo{author}{Begleiter, H.}, \bibinfo{author}{Porjesz, B.}, \bibinfo{author}{Chou, C.}, \bibinfo{author}{Aunon, J.}, \bibinfo{year}{1983}.
\newblock \bibinfo{title}{P3 and stimulus incentive value}.
\newblock \bibinfo{journal}{Psychophysiology} \bibinfo{volume}{20}, \bibinfo{pages}{95--101}.
\bibitem[{Bergstr{\"o}m et~al.(2017)Bergstr{\"o}m, Azevedo, Papiotis, Saldanha and Slater}]{bergstrom2017plausibility}
\bibinfo{author}{Bergstr{\"o}m, I.}, \bibinfo{author}{Azevedo, S.}, \bibinfo{author}{Papiotis, P.}, \bibinfo{author}{Saldanha, N.}, \bibinfo{author}{Slater, M.}, \bibinfo{year}{2017}.
\newblock \bibinfo{title}{The plausibility of a string quartet performance in virtual reality}.
\newblock \bibinfo{journal}{IEEE transactions on visualization and computer graphics} \bibinfo{volume}{23}, \bibinfo{pages}{1352--1359}.
\bibitem[{Brogni et~al.(2003)Brogni, Slater, Steed et~al.}]{brogni2003more}
\bibinfo{author}{Brogni, A.}, \bibinfo{author}{Slater, M.}, \bibinfo{author}{Steed, A.}, et~al., \bibinfo{year}{2003}.
\newblock \bibinfo{title}{More breaks less presence}, in: \bibinfo{booktitle}{Presence 2003: The 6th Annual International Workshop on Presence}, pp. \bibinfo{pages}{1--4}.
\bibitem[{Br{\"u}bach et~al.(2024)Br{\"u}bach, R{\"o}hm, Westermeier, Latoschik and Wienrich}]{brubach2024manipulating}
\bibinfo{author}{Br{\"u}bach, L.}, \bibinfo{author}{R{\"o}hm, M.}, \bibinfo{author}{Westermeier, F.}, \bibinfo{author}{Latoschik, M.E.}, \bibinfo{author}{Wienrich, C.}, \bibinfo{year}{2024}.
\newblock \bibinfo{title}{Manipulating immersion: The impact of perceptual incongruence on perceived plausibility in vr}, in: \bibinfo{booktitle}{2024 IEEE International Symposium on Mixed and Augmented Reality (ISMAR)}, \bibinfo{organization}{IEEE}. pp. \bibinfo{pages}{1078--1086}.
\bibitem[{Br{\"u}bach et~al.(2022)Br{\"u}bach, Westermeier, Wienrich and Latoschik}]{brubach2022breaking}
\bibinfo{author}{Br{\"u}bach, L.}, \bibinfo{author}{Westermeier, F.}, \bibinfo{author}{Wienrich, C.}, \bibinfo{author}{Latoschik, M.E.}, \bibinfo{year}{2022}.
\newblock \bibinfo{title}{Breaking plausibility without breaking presence-evidence for the multi-layer nature of plausibility}.
\newblock \bibinfo{journal}{IEEE Trans Vis Comput Graph} \bibinfo{volume}{28}, \bibinfo{pages}{2267--2276}.
\bibitem[{Brunia et~al.(2011)Brunia, Van~Boxtel and B{\"o}cker}]{brunia2011negative}
\bibinfo{author}{Brunia, C.H.}, \bibinfo{author}{Van~Boxtel, G.J.}, \bibinfo{author}{B{\"o}cker, K.B.}, \bibinfo{year}{2011}.
\newblock \bibinfo{title}{Negative slow waves as indices of anticipation: the bereitschaftspotential, the contingent negative variation, and the stimulus-preceding negativity} .
\bibitem[{Chatrian et~al.(1985)Chatrian, Lettich and Nelson}]{chatrian1985ten}
\bibinfo{author}{Chatrian, G.E.}, \bibinfo{author}{Lettich, E.}, \bibinfo{author}{Nelson, P.L.}, \bibinfo{year}{1985}.
\newblock \bibinfo{title}{Ten percent electrode system for topographic studies of spontaneous and evoked eeg activities}.
\newblock \bibinfo{journal}{American Journal of EEG technology} \bibinfo{volume}{25}, \bibinfo{pages}{83--92}.
\bibitem[{Clark(2013)}]{Clark2013-ah}
\bibinfo{author}{Clark, A.}, \bibinfo{year}{2013}.
\newblock \bibinfo{title}{Whatever next? predictive brains, situated agents, and the future of cognitive science}.
\newblock \bibinfo{journal}{Behav. Brain Sci.} \bibinfo{volume}{36}, \bibinfo{pages}{181--204}.
\bibitem[{Cohn and Kutas(2015)}]{cohn2015getting}
\bibinfo{author}{Cohn, N.}, \bibinfo{author}{Kutas, M.}, \bibinfo{year}{2015}.
\newblock \bibinfo{title}{Getting a cue before getting a clue: Event-related potentials to inference in visual narrative comprehension}.
\newblock \bibinfo{journal}{Neuropsychologia} \bibinfo{volume}{77}, \bibinfo{pages}{267--278}.
\bibitem[{Courchesne et~al.(1975)Courchesne, Hillyard and Galambos}]{courchesne1975stimulus}
\bibinfo{author}{Courchesne, E.}, \bibinfo{author}{Hillyard, S.A.}, \bibinfo{author}{Galambos, R.}, \bibinfo{year}{1975}.
\newblock \bibinfo{title}{Stimulus novelty, task relevance and the visual evoked potential in man}.
\newblock \bibinfo{journal}{Electroencephalography and clinical neurophysiology} \bibinfo{volume}{39}, \bibinfo{pages}{131--143}.
\bibitem[{Cummings and Bailenson(2016)}]{cummings2016immersive}
\bibinfo{author}{Cummings, J.J.}, \bibinfo{author}{Bailenson, J.N.}, \bibinfo{year}{2016}.
\newblock \bibinfo{title}{How immersive is enough? a meta-analysis of the effect of immersive technology on user presence}.
\newblock \bibinfo{journal}{Media psychology} \bibinfo{volume}{19}, \bibinfo{pages}{272--309}.
\bibitem[{Delorme and Makeig(2004)}]{delorme2004eeglab}
\bibinfo{author}{Delorme, A.}, \bibinfo{author}{Makeig, S.}, \bibinfo{year}{2004}.
\newblock \bibinfo{title}{Eeglab: an open source toolbox for analysis of single-trial eeg dynamics including independent component analysis}.
\newblock \bibinfo{journal}{Journal of neuroscience methods} \bibinfo{volume}{134}, \bibinfo{pages}{9--21}.
\bibitem[{Donchin(1981)}]{donchin1981surprise}
\bibinfo{author}{Donchin, E.}, \bibinfo{year}{1981}.
\newblock \bibinfo{title}{Surprise!… surprise?}
\newblock \bibinfo{journal}{Psychophysiology} \bibinfo{volume}{18}, \bibinfo{pages}{493--513}.
\bibitem[{Donchin and Coles(1988)}]{donchin1988p300}
\bibinfo{author}{Donchin, E.}, \bibinfo{author}{Coles, M.G.}, \bibinfo{year}{1988}.
\newblock \bibinfo{title}{Is the p300 component a manifestation of context updating?}
\newblock \bibinfo{journal}{Behavioral and brain sciences} \bibinfo{volume}{11}, \bibinfo{pages}{357--374}.
\bibitem[{Falkenstein(1990)}]{falkenstein1990effects}
\bibinfo{author}{Falkenstein, M.}, \bibinfo{year}{1990}.
\newblock \bibinfo{title}{Effects of errors in choice reaction tasks on the erp under focused and divided attention}.
\newblock \bibinfo{journal}{Psychophysiological brain research} .
\bibitem[{Folstein and Van~Petten(2008)}]{folstein2008influence}
\bibinfo{author}{Folstein, J.R.}, \bibinfo{author}{Van~Petten, C.}, \bibinfo{year}{2008}.
\newblock \bibinfo{title}{Influence of cognitive control and mismatch on the n2 component of the erp: a review}.
\newblock \bibinfo{journal}{Psychophysiology} \bibinfo{volume}{45}, \bibinfo{pages}{152--170}.
\bibitem[{Friston(2010)}]{Friston2010-hy}
\bibinfo{author}{Friston, K.}, \bibinfo{year}{2010}.
\newblock \bibinfo{title}{The free-energy principle: a unified brain theory?}
\newblock \bibinfo{journal}{Nat. Rev. Neurosci.} \bibinfo{volume}{11}, \bibinfo{pages}{127--138}.
\bibitem[{Ganis and Kutas(2003)}]{ganis2003electrophysiological}
\bibinfo{author}{Ganis, G.}, \bibinfo{author}{Kutas, M.}, \bibinfo{year}{2003}.
\newblock \bibinfo{title}{An electrophysiological study of scene effects on object identification}.
\newblock \bibinfo{journal}{Cognitive Brain Research} \bibinfo{volume}{16}, \bibinfo{pages}{123--144}.
\bibitem[{Ganis et~al.(1996)Ganis, Kutas and Sereno}]{ganis1996search}
\bibinfo{author}{Ganis, G.}, \bibinfo{author}{Kutas, M.}, \bibinfo{author}{Sereno, M.I.}, \bibinfo{year}{1996}.
\newblock \bibinfo{title}{The search for “common sense”: An electrophysiological study of the comprehension of words and pictures in reading}.
\newblock \bibinfo{journal}{Journal of cognitive neuroscience} \bibinfo{volume}{8}, \bibinfo{pages}{89--106}.
\bibitem[{Garau et~al.(2008)Garau, Friedman, Widenfeld, Antley, Brogni and Slater}]{garau2008temporal}
\bibinfo{author}{Garau, M.}, \bibinfo{author}{Friedman, D.}, \bibinfo{author}{Widenfeld, H.R.}, \bibinfo{author}{Antley, A.}, \bibinfo{author}{Brogni, A.}, \bibinfo{author}{Slater, M.}, \bibinfo{year}{2008}.
\newblock \bibinfo{title}{Temporal and spatial variations in presence: Qualitative analysis of interviews from an experiment on breaks in presence}.
\newblock \bibinfo{journal}{PRESENCE: Teleoperators and Virtual Environments} \bibinfo{volume}{17}, \bibinfo{pages}{293--309}.
\bibitem[{Gehring et~al.(2007)Gehring, Liu, Orr and Carp}]{gehring2007neural}
\bibinfo{author}{Gehring, W.}, \bibinfo{author}{Liu, Y.}, \bibinfo{author}{Orr, J.}, \bibinfo{author}{Carp, J.}, \bibinfo{year}{2007}.
\newblock \bibinfo{title}{Neural systems for error monitoring: Recent findings and theoretical perspectives}.
\newblock \bibinfo{journal}{International Journal of Psychology} \bibinfo{volume}{43}, \bibinfo{pages}{355}.
\bibitem[{Gehring et~al.(1993)Gehring, Goss, Coles, Meyer and Donchin}]{gehring1993neural}
\bibinfo{author}{Gehring, W.J.}, \bibinfo{author}{Goss, B.}, \bibinfo{author}{Coles, M.G.}, \bibinfo{author}{Meyer, D.E.}, \bibinfo{author}{Donchin, E.}, \bibinfo{year}{1993}.
\newblock \bibinfo{title}{A neural system for error detection and compensation}.
\newblock \bibinfo{journal}{Psychological science} \bibinfo{volume}{4}, \bibinfo{pages}{385--390}.
\bibitem[{Gehrke et~al.(2019)Gehrke, Akman, Lopes, Chen, Singh, Chen, Lin and Gramann}]{gehrke2019detecting}
\bibinfo{author}{Gehrke, L.}, \bibinfo{author}{Akman, S.}, \bibinfo{author}{Lopes, P.}, \bibinfo{author}{Chen, A.}, \bibinfo{author}{Singh, A.K.}, \bibinfo{author}{Chen, H.T.}, \bibinfo{author}{Lin, C.T.}, \bibinfo{author}{Gramann, K.}, \bibinfo{year}{2019}.
\newblock \bibinfo{title}{Detecting visuo-haptic mismatches in virtual reality using the prediction error negativity of event-related brain potentials}, in: \bibinfo{booktitle}{Proceedings of the 2019 CHI conference on human factors in computing systems}, pp. \bibinfo{pages}{1--11}.
\bibitem[{Gehrke et~al.(2022)Gehrke, Lopes, Klug, Akman and Gramann}]{gehrke2022neural}
\bibinfo{author}{Gehrke, L.}, \bibinfo{author}{Lopes, P.}, \bibinfo{author}{Klug, M.}, \bibinfo{author}{Akman, S.}, \bibinfo{author}{Gramann, K.}, \bibinfo{year}{2022}.
\newblock \bibinfo{title}{Neural sources of prediction errors detect unrealistic vr interactions}.
\newblock \bibinfo{journal}{Journal of Neural Engineering} \bibinfo{volume}{19}, \bibinfo{pages}{036002}.
\bibitem[{Gibson(2014)}]{gibson2014ecological}
\bibinfo{author}{Gibson, J.J.}, \bibinfo{year}{2014}.
\newblock \bibinfo{title}{The ecological approach to visual perception: classic edition}.
\newblock \bibinfo{publisher}{Psychology press}.
\bibitem[{Gonzalez-Franco and Lanier(2017)}]{gonzalez2017model}
\bibinfo{author}{Gonzalez-Franco, M.}, \bibinfo{author}{Lanier, J.}, \bibinfo{year}{2017}.
\newblock \bibinfo{title}{Model of illusions and virtual reality}.
\newblock \bibinfo{journal}{Frontiers in psychology} \bibinfo{volume}{8}, \bibinfo{pages}{1125}.
\bibitem[{Gonz{\'a}lez-Franco et~al.(2014)Gonz{\'a}lez-Franco, Peck, Rodr{\'\i}guez-Fornells and Slater}]{gonzalez2014threat}
\bibinfo{author}{Gonz{\'a}lez-Franco, M.}, \bibinfo{author}{Peck, T.C.}, \bibinfo{author}{Rodr{\'\i}guez-Fornells, A.}, \bibinfo{author}{Slater, M.}, \bibinfo{year}{2014}.
\newblock \bibinfo{title}{A threat to a virtual hand elicits motor cortex activation}.
\newblock \bibinfo{journal}{Experimental brain research} \bibinfo{volume}{232}, \bibinfo{pages}{875--887}.
\bibitem[{Gonzalez-Franco et~al.(2010)Gonzalez-Franco, Perez-Marcos, Spanlang and Slater}]{gonzalez2010contribution}
\bibinfo{author}{Gonzalez-Franco, M.}, \bibinfo{author}{Perez-Marcos, D.}, \bibinfo{author}{Spanlang, B.}, \bibinfo{author}{Slater, M.}, \bibinfo{year}{2010}.
\newblock \bibinfo{title}{The contribution of real-time mirror reflections of motor actions on virtual body ownership in an immersive virtual environment}, in: \bibinfo{booktitle}{2010 IEEE virtual reality conference (VR)}, \bibinfo{organization}{IEEE}. pp. \bibinfo{pages}{111--114}.
\bibitem[{Grassini et~al.(2021)Grassini, Laumann, Thorp and Topranin}]{grassini2021using}
\bibinfo{author}{Grassini, S.}, \bibinfo{author}{Laumann, K.}, \bibinfo{author}{Thorp, S.}, \bibinfo{author}{Topranin, V.d.M.}, \bibinfo{year}{2021}.
\newblock \bibinfo{title}{Using electrophysiological measures to evaluate the sense of presence in immersive virtual environments: An event-related potential study}.
\newblock \bibinfo{journal}{Brain and behavior} \bibinfo{volume}{11}, \bibinfo{pages}{e2269}.
\bibitem[{Gratton et~al.(1990)Gratton, Bosco, Kramer, Coles, Wickens and Donchin}]{gratton1990event}
\bibinfo{author}{Gratton, G.}, \bibinfo{author}{Bosco, C.M.}, \bibinfo{author}{Kramer, A.F.}, \bibinfo{author}{Coles, M.G.}, \bibinfo{author}{Wickens, C.D.}, \bibinfo{author}{Donchin, E.}, \bibinfo{year}{1990}.
\newblock \bibinfo{title}{Event-related brain potentials as indices of information extraction and response priming}.
\newblock \bibinfo{journal}{Electroencephalography and clinical neurophysiology} \bibinfo{volume}{75}, \bibinfo{pages}{419--432}.
\bibitem[{Hofer et~al.(2020)Hofer, Hartmann, Eden, Ratan and Hahn}]{hofer2020role}
\bibinfo{author}{Hofer, M.}, \bibinfo{author}{Hartmann, T.}, \bibinfo{author}{Eden, A.}, \bibinfo{author}{Ratan, R.}, \bibinfo{author}{Hahn, L.}, \bibinfo{year}{2020}.
\newblock \bibinfo{title}{The role of plausibility in the experience of spatial presence in virtual environments}.
\newblock \bibinfo{journal}{Frontiers in Virtual Reality} , \bibinfo{pages}{2}.
\bibitem[{Kilteni et~al.(2012)Kilteni, Groten and Slater}]{kilteni2012sense}
\bibinfo{author}{Kilteni, K.}, \bibinfo{author}{Groten, R.}, \bibinfo{author}{Slater, M.}, \bibinfo{year}{2012}.
\newblock \bibinfo{title}{The sense of embodiment in virtual reality}.
\newblock \bibinfo{journal}{Presence: Teleoperators and Virtual Environments} \bibinfo{volume}{21}, \bibinfo{pages}{373--387}.
\bibitem[{Kimura et~al.(2006)Kimura, Katayama and Murohashi}]{kimura2006erp}
\bibinfo{author}{Kimura, M.}, \bibinfo{author}{Katayama, J.}, \bibinfo{author}{Murohashi, H.}, \bibinfo{year}{2006}.
\newblock \bibinfo{title}{An erp study of visual change detection: effects of magnitude of spatial frequency changes on the change-related posterior positivity}.
\newblock \bibinfo{journal}{International Journal of Psychophysiology} \bibinfo{volume}{62}, \bibinfo{pages}{14--23}.
\bibitem[{Kober and Neuper(2012)}]{kober2012using}
\bibinfo{author}{Kober, S.E.}, \bibinfo{author}{Neuper, C.}, \bibinfo{year}{2012}.
\newblock \bibinfo{title}{Using auditory event-related eeg potentials to assess presence in virtual reality}.
\newblock \bibinfo{journal}{International Journal of Human-Computer Studies} \bibinfo{volume}{70}, \bibinfo{pages}{577--587}.
\bibitem[{Kothe et~al.(2024)Kothe, Shirazi, Stenner, Medine, Boulay, Grivich, Mullen, Delorme and Makeig}]{kothe2024lab}
\bibinfo{author}{Kothe, C.}, \bibinfo{author}{Shirazi, S.Y.}, \bibinfo{author}{Stenner, T.}, \bibinfo{author}{Medine, D.}, \bibinfo{author}{Boulay, C.}, \bibinfo{author}{Grivich, M.I.}, \bibinfo{author}{Mullen, T.}, \bibinfo{author}{Delorme, A.}, \bibinfo{author}{Makeig, S.}, \bibinfo{year}{2024}.
\newblock \bibinfo{title}{The lab streaming layer for synchronized multimodal recording}.
\newblock \bibinfo{journal}{BioRxiv} , \bibinfo{pages}{2024--02}.
\bibitem[{Kutas and Federmeier(2011)}]{kutas2011thirty}
\bibinfo{author}{Kutas, M.}, \bibinfo{author}{Federmeier, K.D.}, \bibinfo{year}{2011}.
\newblock \bibinfo{title}{Thirty years and counting: finding meaning in the n400 component of the event-related brain potential (erp)}.
\newblock \bibinfo{journal}{Annual review of psychology} \bibinfo{volume}{62}, \bibinfo{pages}{621--647}.
\bibitem[{Kutas and Hillyard(1980)}]{kutas1980reading}
\bibinfo{author}{Kutas, M.}, \bibinfo{author}{Hillyard, S.A.}, \bibinfo{year}{1980}.
\newblock \bibinfo{title}{Reading senseless sentences: Brain potentials reflect semantic incongruity}.
\newblock \bibinfo{journal}{Science} \bibinfo{volume}{207}, \bibinfo{pages}{203--205}.
\bibitem[{Latoschik and Wienrich(2022)}]{latoschikcongruence}
\bibinfo{author}{Latoschik, M.E.}, \bibinfo{author}{Wienrich, C.}, \bibinfo{year}{2022}.
\newblock \bibinfo{title}{Congruence and plausibility, not presence?! pivotal conditions for xr experiences and effects, a novel approach}.
\newblock \bibinfo{journal}{Frontiers in Virtual Reality} , \bibinfo{pages}{63}.
\bibitem[{Lehser et~al.(2024)Lehser, Hillyard and Strauss}]{lehser2024feeling}
\bibinfo{author}{Lehser, C.}, \bibinfo{author}{Hillyard, S.A.}, \bibinfo{author}{Strauss, D.J.}, \bibinfo{year}{2024}.
\newblock \bibinfo{title}{Feeling senseless sensations: a crossmodal eeg study of mismatched tactile and visual experiences in virtual reality}.
\newblock \bibinfo{journal}{Journal of Neural Engineering} \bibinfo{volume}{21}, \bibinfo{pages}{056042}.
\bibitem[{Lopez-Calderon and Luck(2014)}]{lopez2014erplab}
\bibinfo{author}{Lopez-Calderon, J.}, \bibinfo{author}{Luck, S.J.}, \bibinfo{year}{2014}.
\newblock \bibinfo{title}{Erplab: an open-source toolbox for the analysis of event-related potentials}.
\newblock \bibinfo{journal}{Frontiers in human neuroscience} \bibinfo{volume}{8}, \bibinfo{pages}{213}.
\bibitem[{Makeig et~al.(1997)Makeig, Jung, Bell, Ghahremani and Sejnowski}]{makeig1997blind}
\bibinfo{author}{Makeig, S.}, \bibinfo{author}{Jung, T.P.}, \bibinfo{author}{Bell, A.J.}, \bibinfo{author}{Ghahremani, D.}, \bibinfo{author}{Sejnowski, T.J.}, \bibinfo{year}{1997}.
\newblock \bibinfo{title}{Blind separation of auditory event-related brain responses into independent components}.
\newblock \bibinfo{journal}{Proceedings of the National Academy of Sciences} \bibinfo{volume}{94}, \bibinfo{pages}{10979--10984}.
\bibitem[{Maselli and Slater(2013)}]{maselli2013building}
\bibinfo{author}{Maselli, A.}, \bibinfo{author}{Slater, M.}, \bibinfo{year}{2013}.
\newblock \bibinfo{title}{The building blocks of the full body ownership illusion}.
\newblock \bibinfo{journal}{Frontiers in human neuroscience} \bibinfo{volume}{7}, \bibinfo{pages}{83}.
\bibitem[{Meehan et~al.(2002)Meehan, Insko, Whitton and Brooks~Jr}]{meehan2002physiological}
\bibinfo{author}{Meehan, M.}, \bibinfo{author}{Insko, B.}, \bibinfo{author}{Whitton, M.}, \bibinfo{author}{Brooks~Jr, F.P.}, \bibinfo{year}{2002}.
\newblock \bibinfo{title}{Physiological measures of presence in stressful virtual environments}.
\newblock \bibinfo{journal}{Acm transactions on graphics (tog)} \bibinfo{volume}{21}, \bibinfo{pages}{645--652}.
\bibitem[{Miltner et~al.(1997)Miltner, Braun and Coles}]{miltner1997event}
\bibinfo{author}{Miltner, W.H.}, \bibinfo{author}{Braun, C.H.}, \bibinfo{author}{Coles, M.G.}, \bibinfo{year}{1997}.
\newblock \bibinfo{title}{Event-related brain potentials following incorrect feedback in a time-estimation task: evidence for a “generic” neural system for error detection}.
\newblock \bibinfo{journal}{Journal of cognitive neuroscience} \bibinfo{volume}{9}, \bibinfo{pages}{788--798}.
\bibitem[{N{\"a}{\"a}t{\"a}nen(2001)}]{naatanen2001perception}
\bibinfo{author}{N{\"a}{\"a}t{\"a}nen, R.}, \bibinfo{year}{2001}.
\newblock \bibinfo{title}{The perception of speech sounds by the human brain as reflected by the mismatch negativity (mmn) and its magnetic equivalent (mmnm)}.
\newblock \bibinfo{journal}{Psychophysiology} \bibinfo{volume}{38}, \bibinfo{pages}{1--21}.
\bibitem[{N{\"a}{\"a}t{\"a}nen et~al.(1978)N{\"a}{\"a}t{\"a}nen, Gaillard and M{\"a}ntysalo}]{naatanen1978early}
\bibinfo{author}{N{\"a}{\"a}t{\"a}nen, R.}, \bibinfo{author}{Gaillard, A.W.}, \bibinfo{author}{M{\"a}ntysalo, S.}, \bibinfo{year}{1978}.
\newblock \bibinfo{title}{Early selective-attention effect on evoked potential reinterpreted}.
\newblock \bibinfo{journal}{Acta psychologica} \bibinfo{volume}{42}, \bibinfo{pages}{313--329}.
\bibitem[{Nardi(2024)}]{nardi2024quantifying}
\bibinfo{author}{Nardi, A.}, \bibinfo{year}{2024}.
\newblock \bibinfo{title}{Quantifying breaks in presence in a virtual reality slingshot game using the event-related potential technique}.
\newblock Master's thesis. A. Nardi.
\bibitem[{Neyret et~al.(2020)Neyret, Navarro, Beacco, Oliva, Bourdin, Valenzuela, Barberia and Slater}]{neyret2020embodied}
\bibinfo{author}{Neyret, S.}, \bibinfo{author}{Navarro, X.}, \bibinfo{author}{Beacco, A.}, \bibinfo{author}{Oliva, R.}, \bibinfo{author}{Bourdin, P.}, \bibinfo{author}{Valenzuela, J.}, \bibinfo{author}{Barberia, I.}, \bibinfo{author}{Slater, M.}, \bibinfo{year}{2020}.
\newblock \bibinfo{title}{An embodied perspective as a victim of sexual harassment in virtual reality reduces action conformity in a later milgram obedience scenario}.
\newblock \bibinfo{journal}{Scientific reports} \bibinfo{volume}{10}, \bibinfo{pages}{6207}.
\bibitem[{Nieuwenhuis et~al.(2004)Nieuwenhuis, Holroyd, Mol and Coles}]{nieuwenhuis2004reinforcement}
\bibinfo{author}{Nieuwenhuis, S.}, \bibinfo{author}{Holroyd, C.B.}, \bibinfo{author}{Mol, N.}, \bibinfo{author}{Coles, M.G.}, \bibinfo{year}{2004}.
\newblock \bibinfo{title}{Reinforcement-related brain potentials from medial frontal cortex: origins and functional significance}.
\newblock \bibinfo{journal}{Neuroscience \& Biobehavioral Reviews} \bibinfo{volume}{28}, \bibinfo{pages}{441--448}.
\bibitem[{Nigam et~al.(1992)Nigam, Hoffman and Simons}]{nigam1992n400}
\bibinfo{author}{Nigam, A.}, \bibinfo{author}{Hoffman, J.E.}, \bibinfo{author}{Simons, R.F.}, \bibinfo{year}{1992}.
\newblock \bibinfo{title}{N400 to semantically anomalous pictures and words}.
\newblock \bibinfo{journal}{Journal of cognitive neuroscience} \bibinfo{volume}{4}, \bibinfo{pages}{15--22}.
\bibitem[{No{\"e}(2004)}]{noe2004action}
\bibinfo{author}{No{\"e}, A.}, \bibinfo{year}{2004}.
\newblock \bibinfo{title}{Action in perception}.
\newblock \bibinfo{publisher}{MIT press}.
\bibitem[{Padrao et~al.(2016)Padrao, Gonzalez-Franco, Sanchez-Vives, Slater and Rodriguez-Fornells}]{padrao2016violating}
\bibinfo{author}{Padrao, G.}, \bibinfo{author}{Gonzalez-Franco, M.}, \bibinfo{author}{Sanchez-Vives, M.V.}, \bibinfo{author}{Slater, M.}, \bibinfo{author}{Rodriguez-Fornells, A.}, \bibinfo{year}{2016}.
\newblock \bibinfo{title}{Violating body movement semantics: Neural signatures of self-generated and external-generated errors}.
\newblock \bibinfo{journal}{Neuroimage} \bibinfo{volume}{124}, \bibinfo{pages}{147--156}.
\bibitem[{Pan et~al.(2016)Pan, Slater, Beacco, Navarro, Bellido~Rivas, Swapp, Hale, Forbes, Denvir, de~C.~Hamilton et~al.}]{pan2016responses}
\bibinfo{author}{Pan, X.}, \bibinfo{author}{Slater, M.}, \bibinfo{author}{Beacco, A.}, \bibinfo{author}{Navarro, X.}, \bibinfo{author}{Bellido~Rivas, A.I.}, \bibinfo{author}{Swapp, D.}, \bibinfo{author}{Hale, J.}, \bibinfo{author}{Forbes, P.A.G.}, \bibinfo{author}{Denvir, C.}, \bibinfo{author}{de~C.~Hamilton, A.F.}, et~al., \bibinfo{year}{2016}.
\newblock \bibinfo{title}{The responses of medical general practitioners to unreasonable patient demand for antibiotics-a study of medical ethics using immersive virtual reality}.
\newblock \bibinfo{journal}{PloS one} \bibinfo{volume}{11}, \bibinfo{pages}{e0146837}.
\bibitem[{Peck and Gonzalez-Franco(2021)}]{peck2021avatar}
\bibinfo{author}{Peck, T.C.}, \bibinfo{author}{Gonzalez-Franco, M.}, \bibinfo{year}{2021}.
\newblock \bibinfo{title}{Avatar embodiment. a standardized questionnaire}.
\newblock \bibinfo{journal}{Frontiers in Virtual Reality} \bibinfo{volume}{1}, \bibinfo{pages}{575943}.
\bibitem[{Pion-Tonachini et~al.(2019)Pion-Tonachini, Kreutz-Delgado and Makeig}]{pion2019iclabel}
\bibinfo{author}{Pion-Tonachini, L.}, \bibinfo{author}{Kreutz-Delgado, K.}, \bibinfo{author}{Makeig, S.}, \bibinfo{year}{2019}.
\newblock \bibinfo{title}{Iclabel: An automated electroencephalographic independent component classifier, dataset, and website}.
\newblock \bibinfo{journal}{NeuroImage} \bibinfo{volume}{198}, \bibinfo{pages}{181--197}.
\bibitem[{Polich(2007)}]{polich2007updating}
\bibinfo{author}{Polich, J.}, \bibinfo{year}{2007}.
\newblock \bibinfo{title}{Updating p300: an integrative theory of p3a and p3b}.
\newblock \bibinfo{journal}{Clinical neurophysiology} \bibinfo{volume}{118}, \bibinfo{pages}{2128--2148}.
\bibitem[{Porssut et~al.(2023)Porssut, Iwane, Chavarriaga, Blanke, Mill{\'a}n, Boulic and Herbelin}]{porssut2023eeg}
\bibinfo{author}{Porssut, T.}, \bibinfo{author}{Iwane, F.}, \bibinfo{author}{Chavarriaga, R.}, \bibinfo{author}{Blanke, O.}, \bibinfo{author}{Mill{\'a}n, J.d.R.}, \bibinfo{author}{Boulic, R.}, \bibinfo{author}{Herbelin, B.}, \bibinfo{year}{2023}.
\newblock \bibinfo{title}{Eeg signature of breaks in embodiment in vr}.
\newblock \bibinfo{journal}{Plos one} \bibinfo{volume}{18}, \bibinfo{pages}{e0282967}.
\bibitem[{Pouke et~al.(2022)Pouke, Ylipulli, Uotila, Sitomaniemi, Pouke and Ojala}]{poukePRESENCE}
\bibinfo{author}{Pouke, M.}, \bibinfo{author}{Ylipulli, J.}, \bibinfo{author}{Uotila, E.}, \bibinfo{author}{Sitomaniemi, A.K.}, \bibinfo{author}{Pouke, S.}, \bibinfo{author}{Ojala, T.}, \bibinfo{year}{2022}.
\newblock \bibinfo{title}{{A Qualitative Case Study on Deconstructing Presence for Young Adults and Older Adults}}.
\newblock \bibinfo{journal}{PRESENCE: Virtual and Augmented Reality} \bibinfo{volume}{31}, \bibinfo{pages}{257--281}.
\newblock \DOIprefix\doi{10.1162/pres_a_00397}.
\bibitem[{Raimondo et~al.(2012)Raimondo, Kamienkowski, Sigman and Fernandez~Slezak}]{raimondo2012cudaica}
\bibinfo{author}{Raimondo, F.}, \bibinfo{author}{Kamienkowski, J.E.}, \bibinfo{author}{Sigman, M.}, \bibinfo{author}{Fernandez~Slezak, D.}, \bibinfo{year}{2012}.
\newblock \bibinfo{title}{Cudaica: Gpu optimization of infomax-ica eeg analysis}.
\newblock \bibinfo{journal}{Computational intelligence and neuroscience} \bibinfo{volume}{2012}, \bibinfo{pages}{206972}.
\bibitem[{Rao and Ballard(1999)}]{Rao1999-zr}
\bibinfo{author}{Rao, R.P.}, \bibinfo{author}{Ballard, D.H.}, \bibinfo{year}{1999}.
\newblock \bibinfo{title}{Predictive coding in the visual cortex: a functional interpretation of some extra-classical receptive-field effects}.
\newblock \bibinfo{journal}{Nat. Neurosci.} \bibinfo{volume}{2}, \bibinfo{pages}{79--87}.
\bibitem[{Sanchez-Vives and Slater(2005)}]{sanchez2005presence}
\bibinfo{author}{Sanchez-Vives, M.V.}, \bibinfo{author}{Slater, M.}, \bibinfo{year}{2005}.
\newblock \bibinfo{title}{From presence to consciousness through virtual reality}.
\newblock \bibinfo{journal}{Nature reviews neuroscience} \bibinfo{volume}{6}, \bibinfo{pages}{332--339}.
\bibitem[{Savalle et~al.(2024)Savalle, Pillette, Won, Argelaguet, L{\'e}cuyer and Mac{\'e}}]{savalle2024towards}
\bibinfo{author}{Savalle, E.}, \bibinfo{author}{Pillette, L.}, \bibinfo{author}{Won, K.}, \bibinfo{author}{Argelaguet, F.}, \bibinfo{author}{L{\'e}cuyer, A.}, \bibinfo{author}{Mac{\'e}, M.J.}, \bibinfo{year}{2024}.
\newblock \bibinfo{title}{Towards electrophysiological measurement of presence in virtual reality through auditory oddball stimuli}.
\newblock \bibinfo{journal}{Journal of Neural Engineering} \bibinfo{volume}{21}, \bibinfo{pages}{046015}.
\bibitem[{Seth(2013)}]{Seth2013-jl}
\bibinfo{author}{Seth, A.K.}, \bibinfo{year}{2013}.
\newblock \bibinfo{title}{Interoceptive inference, emotion, and the embodied self}.
\newblock \bibinfo{journal}{Trends Cogn. Sci.} \bibinfo{volume}{17}, \bibinfo{pages}{565--573}.
\bibitem[{Skarbez et~al.(2020)Skarbez, Brooks and Whitton}]{skarbez2020immersion}
\bibinfo{author}{Skarbez, R.}, \bibinfo{author}{Brooks, F.P.}, \bibinfo{author}{Whitton, M.C.}, \bibinfo{year}{2020}.
\newblock \bibinfo{title}{Immersion and coherence: Research agenda and early results}.
\newblock \bibinfo{journal}{IEEE Trans Vis Comput Graph} \bibinfo{volume}{27}, \bibinfo{pages}{3839--3850}.
\bibitem[{Skarbez et~al.(2017a)Skarbez, Brooks and Whitton}]{skarbez2017survey}
\bibinfo{author}{Skarbez, R.}, \bibinfo{author}{Brooks, Jr, F.P.}, \bibinfo{author}{Whitton, M.C.}, \bibinfo{year}{2017}a.
\newblock \bibinfo{title}{A survey of presence and related concepts}.
\newblock \bibinfo{journal}{ACM Computing Surveys (CSUR)} \bibinfo{volume}{50}, \bibinfo{pages}{1--39}.
\bibitem[{Skarbez et~al.(2017b)Skarbez, Neyret, Brooks, Slater and Whitton}]{skarbez2017psychophysical}
\bibinfo{author}{Skarbez, R.}, \bibinfo{author}{Neyret, S.}, \bibinfo{author}{Brooks, F.P.}, \bibinfo{author}{Slater, M.}, \bibinfo{author}{Whitton, M.C.}, \bibinfo{year}{2017}b.
\newblock \bibinfo{title}{A psychophysical experiment regarding components of the plausibility illusion}.
\newblock \bibinfo{journal}{IEEE Trans Vis Comput Graph} \bibinfo{volume}{23}, \bibinfo{pages}{1369--1378}.
\bibitem[{Skarbez(2016)}]{skarbez2016plausibility}
\bibinfo{author}{Skarbez, R.T.}, \bibinfo{year}{2016}.
\newblock \bibinfo{title}{Plausibility illusion in virtual environments}.
\newblock Ph.D. thesis. The University of North Carolina at Chapel Hill.
\bibitem[{Slater(2004)}]{slater2004colorful}
\bibinfo{author}{Slater, M.}, \bibinfo{year}{2004}.
\newblock \bibinfo{title}{How colorful was your day? why questionnaires cannot assess presence in virtual environments}.
\newblock \bibinfo{journal}{Presence} \bibinfo{volume}{13}, \bibinfo{pages}{484--493}.
\bibitem[{Slater(2009)}]{slater2009place}
\bibinfo{author}{Slater, M.}, \bibinfo{year}{2009}.
\newblock \bibinfo{title}{Place illusion and plausibility can lead to realistic behaviour in immersive virtual environments}.
\newblock \bibinfo{journal}{Philosophical Transactions of the Royal Society B: Biological Sciences} \bibinfo{volume}{364}, \bibinfo{pages}{3549--3557}.
\bibitem[{Slater et~al.(2022)Slater, Banakou, Beacco, Gallego, Macia-Varela and Oliva}]{slater2022separate}
\bibinfo{author}{Slater, M.}, \bibinfo{author}{Banakou, D.}, \bibinfo{author}{Beacco, A.}, \bibinfo{author}{Gallego, J.}, \bibinfo{author}{Macia-Varela, F.}, \bibinfo{author}{Oliva, R.}, \bibinfo{year}{2022}.
\newblock \bibinfo{title}{A separate reality: An update on place illusion and plausibility in virtual reality. front}.
\newblock \bibinfo{journal}{Virtual Real. 3: 914392. doi: 10.3389/frvir} .
\bibitem[{Slater et~al.(2023)Slater, Cabriera, Senel, Banakou, Beacco, Oliva and Gallego}]{slater2023sentiment}
\bibinfo{author}{Slater, M.}, \bibinfo{author}{Cabriera, C.}, \bibinfo{author}{Senel, G.}, \bibinfo{author}{Banakou, D.}, \bibinfo{author}{Beacco, A.}, \bibinfo{author}{Oliva, R.}, \bibinfo{author}{Gallego, J.}, \bibinfo{year}{2023}.
\newblock \bibinfo{title}{The sentiment of a virtual rock concert}.
\newblock \bibinfo{journal}{Virtual Reality} \bibinfo{volume}{27}, \bibinfo{pages}{651--675}.
\bibitem[{Slater et~al.(2009)Slater, P{\'e}rez~Marcos, Ehrsson and Sanchez-Vives}]{slater2009inducing}
\bibinfo{author}{Slater, M.}, \bibinfo{author}{P{\'e}rez~Marcos, D.}, \bibinfo{author}{Ehrsson, H.}, \bibinfo{author}{Sanchez-Vives, M.V.}, \bibinfo{year}{2009}.
\newblock \bibinfo{title}{Inducing illusory ownership of a virtual body}.
\newblock \bibinfo{journal}{Frontiers in neuroscience} \bibinfo{volume}{3}, \bibinfo{pages}{29}.
\bibitem[{Slater et~al.(2010)Slater, Spanlang and Corominas}]{slater2010simulating}
\bibinfo{author}{Slater, M.}, \bibinfo{author}{Spanlang, B.}, \bibinfo{author}{Corominas, D.}, \bibinfo{year}{2010}.
\newblock \bibinfo{title}{Simulating virtual environments within virtual environments as the basis for a psychophysics of presence}.
\newblock \bibinfo{journal}{ACM transactions on graphics (TOG)} \bibinfo{volume}{29}, \bibinfo{pages}{1--9}.
\bibitem[{Slater and Steed(2000)}]{slater2000virtual}
\bibinfo{author}{Slater, M.}, \bibinfo{author}{Steed, A.}, \bibinfo{year}{2000}.
\newblock \bibinfo{title}{A virtual presence counter}.
\newblock \bibinfo{journal}{Presence} \bibinfo{volume}{9}, \bibinfo{pages}{413--434}.
\bibitem[{Slater et~al.(1998)Slater, Steed, McCarthy and Maringelli}]{slater1998influence}
\bibinfo{author}{Slater, M.}, \bibinfo{author}{Steed, A.}, \bibinfo{author}{McCarthy, J.}, \bibinfo{author}{Maringelli, F.}, \bibinfo{year}{1998}.
\newblock \bibinfo{title}{The influence of body movement on subjective presence in virtual environments}.
\newblock \bibinfo{journal}{Human factors} \bibinfo{volume}{40}, \bibinfo{pages}{469--477}.
\bibitem[{Slater and Usoh(1993)}]{slater1993representations}
\bibinfo{author}{Slater, M.}, \bibinfo{author}{Usoh, M.}, \bibinfo{year}{1993}.
\newblock \bibinfo{title}{Representations systems, perceptual position, and presence in immersive virtual environments}.
\newblock \bibinfo{journal}{Presence: Teleoperators \& Virtual Environments} \bibinfo{volume}{2}, \bibinfo{pages}{221--233}.
\bibitem[{Slater et~al.(1994)Slater, Usoh and Steed}]{slater1994depth}
\bibinfo{author}{Slater, M.}, \bibinfo{author}{Usoh, M.}, \bibinfo{author}{Steed, A.}, \bibinfo{year}{1994}.
\newblock \bibinfo{title}{Depth of presence in virtual environments}.
\newblock \bibinfo{journal}{PRESENCE: Teleoperators \& Virtual Environments} \bibinfo{volume}{3}, \bibinfo{pages}{130--144}.
\bibitem[{Slater et~al.(1995)Slater, Usoh and Steed}]{slater1995taking}
\bibinfo{author}{Slater, M.}, \bibinfo{author}{Usoh, M.}, \bibinfo{author}{Steed, A.}, \bibinfo{year}{1995}.
\newblock \bibinfo{title}{Taking steps: the influence of a walking technique on presence in virtual reality}.
\newblock \bibinfo{journal}{ACM Transactions on Computer-Human Interaction (TOCHI)} \bibinfo{volume}{2}, \bibinfo{pages}{201--219}.
\bibitem[{Slater and Wilbur(1997)}]{slater1997framework}
\bibinfo{author}{Slater, M.}, \bibinfo{author}{Wilbur, S.}, \bibinfo{year}{1997}.
\newblock \bibinfo{title}{A framework for immersive virtual environments (five): Speculations on the role of presence in virtual environments}.
\newblock \bibinfo{journal}{PRESENCE: Teleoperators \& Virtual Environments} \bibinfo{volume}{6}, \bibinfo{pages}{603--616}.
\bibitem[{Suzuki et~al.(2023)Suzuki, Mariola, Schwartzman and Seth}]{suzuki2023using}
\bibinfo{author}{Suzuki, K.}, \bibinfo{author}{Mariola, A.}, \bibinfo{author}{Schwartzman, D.J.}, \bibinfo{author}{Seth, A.K.}, \bibinfo{year}{2023}.
\newblock \bibinfo{title}{Using extended reality to study the experience of presence}, in: \bibinfo{booktitle}{Virtual Reality in Behavioral Neuroscience: New Insights and Methods}. \bibinfo{publisher}{Springer}, pp. \bibinfo{pages}{255--285}.
\bibitem[{Tromp et~al.(2018)Tromp, Peeters, Meyer and Hagoort}]{tromp2018combined}
\bibinfo{author}{Tromp, J.}, \bibinfo{author}{Peeters, D.}, \bibinfo{author}{Meyer, A.S.}, \bibinfo{author}{Hagoort, P.}, \bibinfo{year}{2018}.
\newblock \bibinfo{title}{The combined use of virtual reality and eeg to study language processing in naturalistic environments}.
\newblock \bibinfo{journal}{Behavior research methods} \bibinfo{volume}{50}, \bibinfo{pages}{862--869}.
\bibitem[{Twomey et~al.(2015)Twomey, Murphy, Kelly and O'Connell}]{twomey2015classic}
\bibinfo{author}{Twomey, D.M.}, \bibinfo{author}{Murphy, P.R.}, \bibinfo{author}{Kelly, S.P.}, \bibinfo{author}{O'Connell, R.G.}, \bibinfo{year}{2015}.
\newblock \bibinfo{title}{The classic p300 encodes a build-to-threshold decision variable}.
\newblock \bibinfo{journal}{European journal of neuroscience} \bibinfo{volume}{42}, \bibinfo{pages}{1636--1643}.
\bibitem[{Usoh et~al.(2000)Usoh, Catena, Arman and Slater}]{usoh2000using}
\bibinfo{author}{Usoh, M.}, \bibinfo{author}{Catena, E.}, \bibinfo{author}{Arman, S.}, \bibinfo{author}{Slater, M.}, \bibinfo{year}{2000}.
\newblock \bibinfo{title}{Using presence questionnaires in reality}.
\newblock \bibinfo{journal}{PRESENCE} \bibinfo{volume}{9}, \bibinfo{pages}{497--503}.
\bibitem[{Van Der~Hoort and Ehrsson(2016)}]{van2016illusions}
\bibinfo{author}{Van Der~Hoort, B.}, \bibinfo{author}{Ehrsson, H.H.}, \bibinfo{year}{2016}.
\newblock \bibinfo{title}{Illusions of having small or large invisible bodies influence visual perception of object size}.
\newblock \bibinfo{journal}{Scientific reports} \bibinfo{volume}{6}, \bibinfo{pages}{1--9}.
\bibitem[{Van Der~Hoort et~al.(2011)Van Der~Hoort, Guterstam and Ehrsson}]{van2011being}
\bibinfo{author}{Van Der~Hoort, B.}, \bibinfo{author}{Guterstam, A.}, \bibinfo{author}{Ehrsson, H.H.}, \bibinfo{year}{2011}.
\newblock \bibinfo{title}{Being barbie: the size of one’s own body determines the perceived size of the world}.
\newblock \bibinfo{journal}{PloS one} \bibinfo{volume}{6}, \bibinfo{pages}{e20195}.
\bibitem[{Vasconcelos-Raposo et~al.(2016)Vasconcelos-Raposo, Bessa, Melo, Barbosa, Rodrigues, Teixeira, Cabral and Sousa}]{vasconcelos2016adaptation}
\bibinfo{author}{Vasconcelos-Raposo, J.}, \bibinfo{author}{Bessa, M.}, \bibinfo{author}{Melo, M.}, \bibinfo{author}{Barbosa, L.}, \bibinfo{author}{Rodrigues, R.}, \bibinfo{author}{Teixeira, C.M.}, \bibinfo{author}{Cabral, L.}, \bibinfo{author}{Sousa, A.A.}, \bibinfo{year}{2016}.
\newblock \bibinfo{title}{Adaptation and validation of the igroup presence questionnaire (ipq) in a portuguese sample}.
\newblock \bibinfo{journal}{Presence} \bibinfo{volume}{25}, \bibinfo{pages}{191--203}.
\bibitem[{V{\~o} and Wolfe(2013)}]{vo2013differential}
\bibinfo{author}{V{\~o}, M.L.H.}, \bibinfo{author}{Wolfe, J.M.}, \bibinfo{year}{2013}.
\newblock \bibinfo{title}{Differential electrophysiological signatures of semantic and syntactic scene processing}.
\newblock \bibinfo{journal}{Psychological science} \bibinfo{volume}{24}, \bibinfo{pages}{1816--1823}.
\bibitem[{Voeten and Voeten(2021)}]{voeten2021package}
\bibinfo{author}{Voeten, C.C.}, \bibinfo{author}{Voeten, M.C.C.}, \bibinfo{year}{2021}.
\newblock \bibinfo{title}{Package ‘buildmer’}.
\bibitem[{Wickens et~al.(1983)Wickens, Kramer, Vanasse and Donchin}]{wickens1983performance}
\bibinfo{author}{Wickens, C.}, \bibinfo{author}{Kramer, A.}, \bibinfo{author}{Vanasse, L.}, \bibinfo{author}{Donchin, E.}, \bibinfo{year}{1983}.
\newblock \bibinfo{title}{Performance of concurrent tasks: a psychophysiological analysis of the reciprocity of information-processing resources}.
\newblock \bibinfo{journal}{Science} \bibinfo{volume}{221}, \bibinfo{pages}{1080--1082}.
\bibitem[{Witmer and Singer(1998)}]{witmer1998measuring}
\bibinfo{author}{Witmer, B.G.}, \bibinfo{author}{Singer, M.J.}, \bibinfo{year}{1998}.
\newblock \bibinfo{title}{Measuring presence in virtual environments: A presence questionnaire}.
\newblock \bibinfo{journal}{Presence} \bibinfo{volume}{7}, \bibinfo{pages}{225--240}.
\bibitem[{Zhang et~al.(2024)Zhang, Garrett and Luck}]{zhang2024optimal}
\bibinfo{author}{Zhang, G.}, \bibinfo{author}{Garrett, D.R.}, \bibinfo{author}{Luck, S.J.}, \bibinfo{year}{2024}.
\newblock \bibinfo{title}{Optimal filters for erp research ii: Recommended settings for seven common erp components}.
\newblock \bibinfo{journal}{Psychophysiology} \bibinfo{volume}{61}, \bibinfo{pages}{e14530}.
\bibitem[{Zimmons and Panter(2003)}]{zimmons2003influence}
\bibinfo{author}{Zimmons, P.}, \bibinfo{author}{Panter, A.}, \bibinfo{year}{2003}.
\newblock \bibinfo{title}{The influence of rendering quality on presence and task performance in a virtual environment}, in: \bibinfo{booktitle}{IEEE Virtual Reality, 2003. Proceedings.}, \bibinfo{organization}{IEEE}. pp. \bibinfo{pages}{293--294}.

\end{thebibliography}






\end{document}